\documentclass[]{aastex631}

\usepackage{longtable}
\usepackage{float}
\newcommand{\angstrom}{\text{\normalfont\AA}}
\usepackage{amsfonts}
\usepackage{amsmath}
\usepackage{amssymb}
\usepackage{mathrsfs}

\shorttitle{The three regimes of atmospheric evaporation}
\shortauthors{Modirrousta-Galian \& Korenaga}

\graphicspath{{./}{figures/}}

\begin{document}

\title{The three regimes of atmospheric evaporation for super-Earths and sub-Neptunes}

\author[0000-0001-6425-9415]{Darius Modirrousta-Galian}
\affiliation{Yale University \\
Department of Earth and Planetary Sciences,\\ 
210 Whitney Ave., New Haven, CT 06511, USA}

\author[0000-0002-4785-2273]{Jun Korenaga}
\affiliation{Yale University \\
Department of Earth and Planetary Sciences,\\ 
210 Whitney Ave., New Haven, CT 06511, USA}

\begin{abstract}
    A significant fraction of super-Earths and sub-Neptunes are thought to experience an extreme loss of volatiles because of atmospheric evaporation in the early stages of their life. Though the mechanisms behind the extreme mass loss are not fully understood, two contenders have been widely discussed: photoevaporation from X-ray and ultraviolet irradiation and core powered mass loss. Here, it is shown that both mechanisms occur but with different timescales, and that atmospheric loss can take place over three regimes. In the first regime, a planet has very high internal temperatures arising from its high-energy formation processes. These high temperatures give rise to a fully convecting atmosphere that efficiently loses mass without much internal cooling. The second regime applies to planets with lower internal temperatures, so a radiative region forms but the photosphere still remains outside the Bondi radius. Hence, mass loss continues to depend only on the internal temperatures. Planets with the lowest internal temperatures are in the third regime, when the photosphere forms below the Bondi radius and mass is lost primarily because of X-ray and ultraviolet irradiation. This paper provides the first unifying framework for modeling atmospheric evaporation through the lifespan of a planet.
\end{abstract}
 

\section{Introduction} 
\label{sec:intro}

Atmospheric evaporation is the process through which gases are lost from planetary atmospheres. This can occur through a variety of mechanisms, including thermal effects \citep[e.g.,][]{Jeans1925,Spitzer1949,Chamberlain1962}, mechanical impact erosion \citep[e.g.,][]{Cameron1983,Ahrens1993, Genda2005,Schlichting2015}, coronal mass ejections \citep[e.g.,][]{Cohen2011,Hazra2022}, X-ray and ultraviolet (XUV) irradiation \citep[e.g.,][]{Watson1981, Kasting1983, Yelle2004, Garcia2007,Tian2015, Kubyshkina2018(1),Modirrousta2020b}, and high interior temperatures \citep{Gupta2018,Biersteker2019,Biersteker2021}. Modeling suggests that the last two mechanisms are most dominant because of the large amount of available energy to drive atmospheric outflow \citep[e.g.,][]{Ginzburg2016,Micela2022}. Most studies investigating the effects of atmospheric evaporation have employed idealized conditions, such as only XUV irradiation \citep[e.g.,][]{Erkaev2007,Murray2009,Owen2013} or only interior energy influencing mass loss \citep[e.g.,][]{Gupta2018,Biersteker2019,Biersteker2021}. However, these mechanisms are not mutually exclusive and can trigger atmospheric outflow within their respective timescales. Evaluating the contribution of each mechanism requires a comprehensive and self-consistent model for atmospheric evaporation. The objective of this paper is to build such a model so that the evolutions and histories of super-Earth and sub-Neptune exoplanets can be constrained. We focus only on super-Earths and sub-Neptunes because their atmospheres constitute a small fraction of their total mass \citep[e.g.,][]{Ikoma2012,Lopez2014}, so their geophysical and atmospheric conditions need to be considered concurrently when modeling mass loss. Larger bodies, like gas giants, require different physics because they are composed mostly of hydrogen, and thus the significance of geophysical conditions is unclear \citep[e.g.,][]{Helled2017}.

In this paper, we model the rate of atmospheric evaporation of exoplanets as a function of their properties, such as their surface and equilibrium temperatures, as well as the XUV flux they are exposed to. It will be shown that atmospheric evaporation can be categorized under three regimes. Regime one applies to the hottest exoplanets with, as suggested for Earth \citep[e.g.,][]{Cameron1991,Cameron1997,Canup2008,Karato2014,Nakajima2015,Lock2018}, very high surface temperatures immediately after formation ($>$10,000~K). Under such conditions, a primordial atmosphere can be fully convecting from its base to the Bondi radius \citep[i.e., the radius at which the atmosphere is no longer gravitationally bound;][]{Ginzburg2016} if it satisfies a minimum mass requirement (see section~\ref{sec:regime_one_requirements}). Regime two applies to planets with lower internal temperatures when a radiative region forms at the top of the atmosphere, but the photosphere (the section of the planet's atmosphere where the optical depth of thermal photons, $\tau_{\rm th}$, is 2/3) is still located outside the Bondi radius. The entire atmosphere is therefore optically thick, so X-ray and ultraviolet photons have no influence on mass loss (see section~\ref{sec:regime_3}). Planets with even lower internal temperatures are in regime three, during which the photosphere is located beneath the Bondi radius, and gas between the Bondi radius and the photosphere is optically thin. XUV photons will therefore become a major source of heating that will lead to photoevaporation. 
In our model, we assume a constant atmospheric composition and therefore do not include processes such as volcanism \citep[e.g.,][]{Orourke2012}, late accretion \citep[e.g.,][]{Marchi2018}, or the entrainment and subsequent mixing of surface materials in the atmosphere \citep[e.g.,][]{Moll2017}. Tidal and centrifugal effects \citep[e.g.,][]{Modirrousta2020} are also not included. This treatment is sufficient for our purposes because it allows us to isolate and constrain the loss of primordial gases from super-Earth and sub-Neptune exoplanets owing to internal energy and stellar irradiation. Further complexity can be incorporated in future studies. In what follows, we first describe the terms and parameters used in our model. We then describe our theory, summarize our results, and compare our findings with other models in the literature. This paper concludes with a summary of our findings.

\section{Nomenclature}

The following isentropic relations for an ideal gas will be assumed for the temperature profile of a convecting gas:
\begin{equation}
    \frac{T_{2}}{T_{1}} = \left(\frac{P_{2}}{P_{1}}\right)^{\frac{\gamma-1}{\gamma}} = \left(\frac{\rho_{2}}{\rho_{1}}\right)^{{\gamma-1}},
\label{eq:isentropic_relations}
\end{equation}
where $T$ is the temperature, $P$ is the pressure, $\rho$ is the density, and $\gamma$ is the heat capacity ratio $c_{\rm p}/c_{\rm v}$. Equation~\ref{eq:isentropic_relations} assumes a calorically perfect gas, where $\gamma$ is constant. In real systems, however, $\gamma$ is not constant because it changes with temperature, pressure, and the composition of the gas \citep[e.g.,][]{Burm1999,Capitelli2008,Capitelli2009}. Indeed, a more accurate prescription may be incorporated by, for example, adopting the equation of state of a hydrogen-helium mixture \citep[e.g.,][]{Militzer2013,Becker2014}, as well as including the effects of condensation \citep[e.g.,][]{Nakajima1992}. Whereas such additions provide a more realistic description of planetary physics, they do not negate the main point of this study, that is, mass loss models must reflect the evolving thermodynamic conditions of planetary interiors. Furthermore, large uncertainties in observable features (e.g., mass, radius, and atmospheric composition), together with our rudimentary understanding of exoplanet interiors, imply that the precise parameter choices used in this paper matter little. To this end, our theoretical framework focuses on unifying geodynamical and atmospheric principles in the context of atmospheric evaporation, constituting a starting point for more thorough treatments in future studies.

When discussing atmospheric evaporation, it is necessary to define the top and bottom boundaries of the atmosphere. The bottom boundary is given by the interface between the atmosphere and the surface of the condensed section of the planet, whereas the upper boundary is the smallest of the Bondi radius, the Hill sphere, and the exobase. After the protoplanetary disk has dissipated, the Bondi radius is given by
\begin{equation}
    R_{\rm B} = \frac{2GM_{\rm n}\mu_{\rm B}}{\gamma k_{\rm B}T_{\rm B}},
\label{eq:Bondi_radius}
\end{equation}
with $G$, $M_{\rm n}$, $k_{\rm B}$, $\mu_{\rm B}$, and $T_{\rm B}$ being the gravitational constant, planetary mass, Boltzmann's constant, and mean molecular mass and temperature at the Bondi radius, respectively. A more complex formulation is required if the protoplanetary disk is extant because the velocity of the planet and gaseous environment need to be considered concurrently \citep[e.g.,][]{Armitage2014}. In this study, we begin tracking the rate of atmospheric evaporation after the protoplanetary disk has dissipated, so equation~\ref{eq:Bondi_radius} applies. The Bondi radius is defined as the location where the average gas particle achieves escape velocity and is therefore no longer gravitationally bound \citep[e.g.,][]{Ginzburg2016}. An alternative name for this radius is the sonic point \citep[e.g.,][]{Parker1964}. The Hill sphere (also known as the Roche lobe radius or the first and second Lagrange points) is defined as the region in space where an astronomical body is gravitationally dominant,
\begin{equation}
    R_{\rm Hi} = a\left(\frac{M_{\rm n}}{3M_{\ast}}\right)^{1/3},
\label{eq:Hill_sphere}
\end{equation}
where $a$ and $M_{\ast}$ are the semi-major axis and stellar mass, respectively. Equation~\ref{eq:Hill_sphere} applies only to a planet with a significantly smaller mass than its host star and with a near-circular orbit. The exobase is defined as the location where the mean free path of a particle is greater than the local scale height, so gases beyond this location become collisionless and no longer follow the ideal gas equation. This radius is given by
\begin{equation}
    R_{\rm x} = \left(\frac{GM_{\rm n}\mu_{\rm x}^{2}}{\sqrt{2}\pi d^{2} \rho_{\rm x} k_{\rm B}T_{\rm x}}\right)^{\frac{1}{2}},
\end{equation}
where $T_{x}$ is the exobase temperature that is given by the models of \citet{Bates1951,Bates1959}, $d$ is the mean kinetic diameter of local particles that is defined as $(\sqrt{2}\pi l n)^{-1/2}$ for a pure gas, with $l$ being the mean free path of a particle, and $n$ is the particle number density. The kinetic diameter is larger than the atomic diameter, which is defined by the electron shell, because it describes the area of influence of the particle. 

The Bondi radius, Hill sphere, and exobase all depend on the planetary mass, which we assume is well approximated by the mass of the central condensed section of the planet. We make this assumption because formation models \citep[e.g.,][]{Ikoma2012} and atmospheric models \citep[e.g.,][]{Lopez2014} suggest that most super-Earths and sub-Neptunes form with atmospheres that are ${\sim}$1\% of their total planetary masses. We therefore consider Earth-like planets with primordial atmospheres that have a negligible mass compared to the central condensed part. It is common to refer to this central section as the core in the exoplanetary literature, but in Earth and planetary sciences, this term refers to the metallic section beneath the silicate mantle of a planet. To avoid confusion, we call the central condensed part the planetary nucleus whereas the metallic center of this nucleus is referred to as the metallic core. The total planetary mass and the mass of the nucleus will be used interchangeably because we focus on planets with primordial atmospheres that have a negligible mass compared to the mass of the nucleus.

Beyond the exobase, a gas becomes collisionless and can no longer accelerate during its adiabatic expansion. If the exobase were therefore smaller than $R_{\rm B}$ and $R_{\rm Hi}$, the atmosphere would not be hydrodynamic and mass loss would occur through the inefficient process of Jeans escape (see appendix for derivation)
\begin{equation}
    \dot{M}_{\rm a} = \frac{2GM\mu_{\rm x}}{d^{2}}\left(\frac{\mu_{\rm x}}{\pi k_{\rm B}T_{\rm x}}\right)^{\frac{1}{2}}\left(\frac{GM_{\rm n}\mu_{\rm x}}{k_{\rm B}T_{\rm x}R_{\rm x}}+1\right)\exp{\left(-\frac{GM_{\rm n}\mu_{\rm x}}{k_{\rm B}T_{\rm x}R_{\rm x}}\right)}.
\end{equation}
The planets of interest to this study have primordial atmospheres and are warm or hot, so they undergo hydrodynamic mass loss. The exosphere must therefore be larger than $R_{\rm B}$ or $R_{\rm Hi}$, so $R_{\rm x}$ is not relevant to this study and will thus not be further discussed. Combining equation~\ref{eq:Bondi_radius} with equation~\ref{eq:Hill_sphere} and solving for the Bondi temperature gives the temperature condition for the Bondi radius to be smaller than the Hill sphere:
\begin{equation}
    T_{\rm B} > 48\left(\frac{\gamma}{\gamma_{0}}\right)^{-1}\left(\frac{M_{\rm n}}{M_{\oplus}}\right)^{\frac{2}{3}} \left(\frac{M_{\ast}}{M_{\odot}}\right)^{\frac{1}{3}} \left(\frac{\mu}{\rm amu}\right) \left(\frac{a}{\rm AU}\right)^{-1},
\label{eq:temperature_condition}
\end{equation}
where $\gamma_{0}$ (with a value of 4/3) is the reference heat capacity ratio. The planets of interest to this study have high Bondi temperatures because of their large interior luminosities and the high XUV irradiation they are exposed to, so equation~\ref{eq:temperature_condition} is generally satisfied. We will therefore use the Bondi radius as the upper limit for the atmosphere.

Throughout this document, we make use of the optical depth (also known as optical thickness) that can be expressed as a function of the opacity of the medium $d\tau {=} \rho(r) \kappa dr$, where $\rho$ is the density and $\kappa$ is the opacity. The optical depth applies to any wavelength and not just optical photons ($\rm 380{-}700~nm$). In this study, we distinguish between the optical depths of thermal and XUV photons. Gases usually have wavelength-dependent opacities, so a region where a gas is optically thick to photons of a particular wavelength, such as XUV photons ($\tau_{\rm XUV} {\geq} 2/3 $), may not be optically thick to photons of another wavelength, like thermal photons ($\tau_{\rm th} {<} 2/3 $). Tables~\ref{tab:subscripts} and \ref{tab:params} in the appendix summarize the terms, parameters, and constants used in this paper. All equations, terms, parameters, and values use the International System of Units (SI).

\section{Overview and general setup}
\label{sec:general_model_setup}

\begin{figure}[ht]
    \centering
    \includegraphics[width=\textwidth]{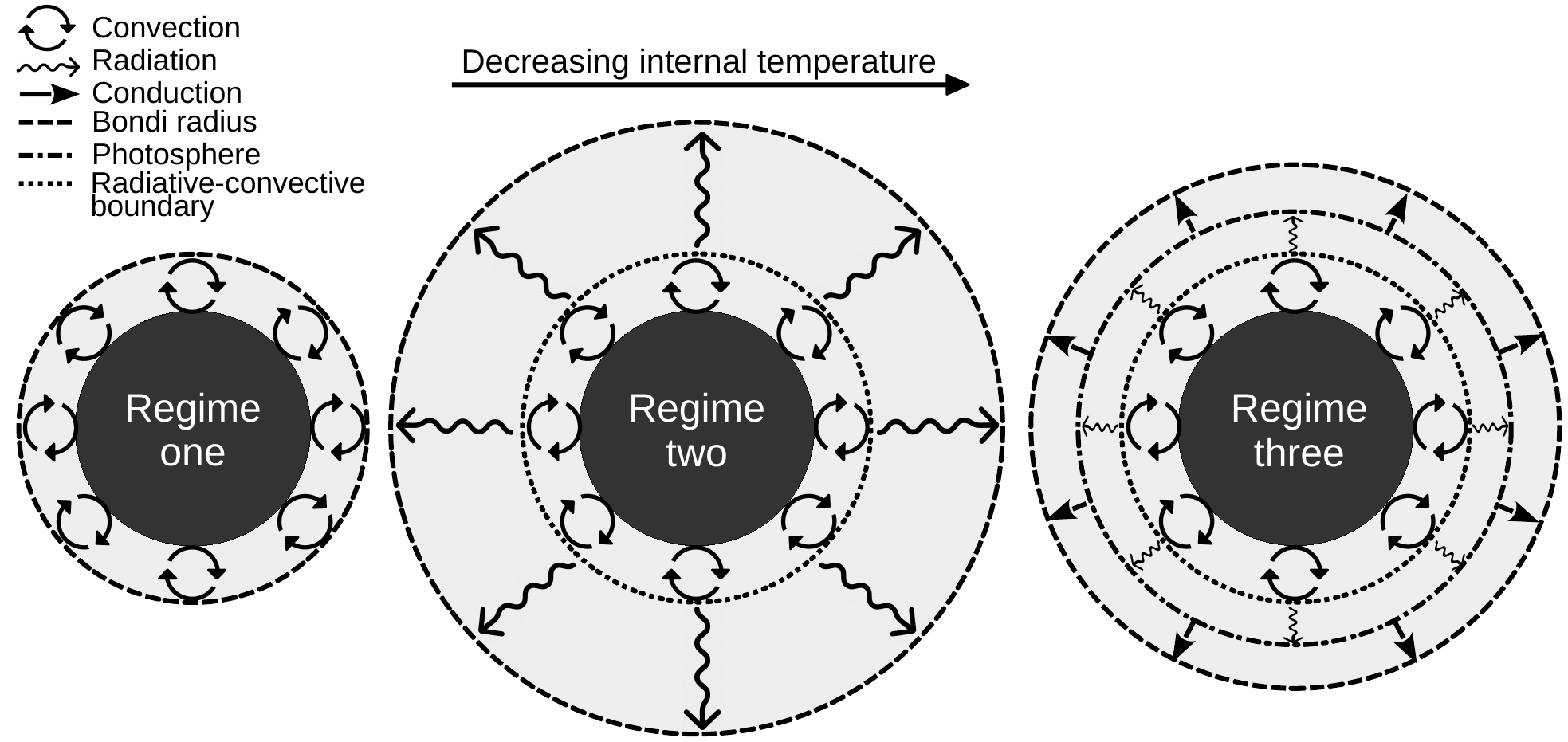}
    \caption{Schematic diagram showing planetary nuclei with overlying atmospheres exhibiting the three regimes of atmospheric evaporation. Regime one is for a fully convecting atmosphere. Regime two is for a convecting and radiative atmosphere. Regime three is for a convecting, radiative, and conducting atmosphere. Internal temperatures decrease from left to right. Diagram not to scale.}
    \label{fig:atmospheric_cartoon}
\end{figure}

In this section, we outline our findings and include the general assumptions underlying our theoretical framework. Each of the three regimes described in our model corresponds to different internal thermodynamic conditions, with regime one being the hottest and regime three the coldest (see figure~\ref{fig:atmospheric_cartoon}). The radiative-convective boundary lies outside the Bondi radius in regime one, so the atmosphere is fully convecting. In regime two, the radiative-convective boundary is below the Bondi radius so that radiation is the most efficient energy transfer mechanism in the upper sections of the atmosphere. A thermosphere has still not formed because of the high internal temperatures, so X-ray and ultraviolet irradiation do not contribute to atmospheric evaporation. A planet enters into regime three at even lower temperatures, where it develops a thermosphere in which conduction is the main source of energy transport. A full description of each regime can be found in sections~\ref{sec:regime_1} to \ref{sec:regime_3}.

The internal energy of a planet comes from four sources: (1) radiogenic heating, (2) tidal heating, (3) the accretionary energy of the primordial atmosphere, and (4) the accretionary energy of the nucleus \citep[for a review on planetary heat sources, see][]{Hussmann2010}. Each of the above mentioned mechanisms acts on a different timescale, which depends on the configuration of the system. The conditions immediately after formation are most relevant to this study because it is when a planet is hottest and the host star is most active, leading to the highest mass loss. It can be shown that the first three mechanisms are less significant by evaluating their magnitudes in the initial stages of a planet's life.

Using a galactic chemical evolution model, \citet{Frank2014} estimated the typical radiogenic heating of Earth-like exoplanets. According to their estimate, the heat flow from the decay of $^{40}$K, $^{232}$Th, $^{235}$U, and $^{238}$U is at most $\rm 2~W~m^{-2}$, which is sufficient to raise surface temperatures by only one hundred Kelvin. Tidal heating is sensitive to the initial conditions of the planet, such as its eccentricity, orbital distance, and the mass of the star and planet. \citet{Jackson2008(1)} suggested that heat fluxes rarely exceed $\rm 1000~W~m^{-2}$, corresponding to surface temperature of two thousand kelvin. The accretionary energy of the atmosphere is minor compared to that of the central nucleus because the atmospheres of super-Earths and sub-Neptunes constitute only a small percentage of the total planetary mass \citep{Ikoma2012,Lopez2014}. Regarding the accretionary energy of the nucleus, planets with masses comparable to Earth and greater are thought to undergo a giant impact phase in which they collide and fuse with large planetary embryos, releasing massive amounts of energy in the process \citep[e.g.,][]{Benz1986,Kokubo1998,Kokubo2000,Elser2011}. Simulations suggest that initial surface temperatures can exceed 10,000~K \citep[e.g.,][]{Cameron1991,Cameron1997,Canup2008,Karato2014,Nakajima2015,Lock2018}, which is significantly higher than what is achievable with the aforementioned mechanisms. We therefore focus only on the accretionary energy of the central nucleus. Whereas our framework does not explicitly include the mechanical effects of impact-induced atmospheric loss, simulations suggest that only a small fraction (${\sim}10\%$) of the atmosphere would be removed \citep[e.g.,][]{Cameron1983,Ahrens1993,Genda2005,Schlichting2015}, and the effect of this removal on the subsequent evolution can easily be evaluated by simply reducing the atmospheric mass. We therefore assume that the mechanical effects of the impact are quenched in a geologically negligible timescale, and equilibrium conditions apply from the beginning of our simulation. We will treat the initial surface temperature as a free parameter because the accretionary energy of the nucleus depends on the history of planetary accretion. 

In our initial model setup, we consider a planet with an entirely molten mantle and an optically thick gravitationaly bound primordial atmosphere (see Figure~\ref{fig:regime_one}). Our objective is to build a general atmospheric evaporation model instead of analyzing a planet or a group of planets in specific, so we leave the planetary properties as free parameters. Because atmospheric evaporation is dictated by the conditions at the Bondi radius, the Bondi density ($\rho_{\rm B}$) and temperature ($T_{\rm B}$) will be found to evaluate the mass loss rate. The other bulk parameters, such as the mass of the nucleus, atmospheric mass, and equilibrium temperature are assumed to be known. Because the cooling rate can only be quantified by knowing the conditions of the magma ocean, we provide a prescription for modeling secular cooling in the following section.

\subsection{Modeling secular cooling}

Our simulation starts with a planet in a very hot initial state that cools gradually. It is necessary to model the top boundary layer of the magma ocean, the bottom boundary layer of the atmosphere, and the optical depth of the atmosphere to calculate the cooling rate. In the following sections, we describe our numerical prescription for modeling the boundary layers and optical depth of the atmosphere.

\subsubsection{Magma ocean boundary layer}

The top boundary layer of the magma ocean is the means through which the planetary nucleus cools. We adopt the classical Rayleigh–Bénard convection scaling of $\rm Nu{\propto} Ra^{1/3}$ in this paper, with $\rm Nu$ and $\rm Ra$ being the Nusselt and Rayleigh numbers, respectively \citep{Priestley1954,Malkus1954,Howard1966}. The heat flux through the magma ocean boundary layer can be parameterized as \citep{Solomatov2015},
\begin{equation}
    F_{\rm s} = 0.089\left[k_{\rm mb}\rho_{\rm mb}\left(T_{\rm s}-T_{\rm mb}\right)^{2}\right]^{\frac{2}{3}}\left(\frac{c_{\rm p,mb}g_{\rm mb}\alpha_{\rm mb}}{\eta_{\rm mb}}\right)^{\frac{1}{3}},
\label{eq:flux_through_magma}
\end{equation}
where $k_{\rm mb}$ is the thermal conductivity \citep[$\rm {\sim}2~W~K^{-1}~m^{-1}$;][]{Lesher2015}, $\rho_{\rm mb}$ is the density \citep[$\rm {\sim}4000~kg~m^{-3}$;][]{Solomatov2015}, $T_{\rm s}$ is the surface temperature of the magma ocean, $T_{\rm mb}$ is the temperature at the bottom of the surface boundary layer of the magma ocean, $g_{\rm mb}$ is the gravitational acceleration, $\alpha_{\rm mb}$ is the volumetric thermal expansion coefficient \citep[$\rm {\sim}5{\times}10^{-5}~K^{-1}$;][]{Solomatov2015}, and $\eta_{\rm mb}$ is the dynamic viscosity \citep[$\rm {\sim}0.1~Pa~s$;][]{Solomatov2015}. For the specific heat of the mantle, we adopt a constant value of $c_{\rm p,mb} {=} 5000~{\rm J~K~kg^{-1}}$, which includes pressure and temperature effects, as well as the enthalpy of crystallization \citep{Miyazaki2019b}. The energy balance equation is therefore
\begin{equation}
    M_{\rm mo}c_{\rm p,mb}\frac{dT_{\rm mb}}{dt} = 4\pi R_{\rm s}^{2} F_{\rm s},
\label{eq:interior_cooling}
\end{equation} 
where the heat flux from the metallic core is ignored because it is generally small \citep{Stevenson1983,Orourke2017}. In the context of atmospheric evaporation, $T_{\rm mb}$ is the most fundamental parameter of the planet's internal state because it is a measure of the available energy to drive outflow. The temperature at the bottom of the convecting section of the atmosphere, $T_{\rm ab}$, is found by determining the temperature contrast across the magma and atmospheric boundary layers. Because the flux through the boundaries is not known a priori, the temperature contrasts are calculated through iteration. After $T_{\rm ab}$ is known, the conditions at the Bondi radius can be determined through the relevant model for the regime the atmosphere is in.

\subsubsection{Atmospheric boundary layer}

The conditions of the thin atmospheric boundary layer above the surface of the magma ocean are described by
\begin{equation}
    {\rm Ra}_{\rm cr} = \frac{\rho^{2} c_{\rm p} g \alpha \left(T_{\rm ab}-T_{\rm s} \right) \left(R_{\rm ab}-R_{\rm s}\right)^{3}}{\eta k}
\label{eq:critical_Rayleigh_number}
\end{equation}
and
\begin{equation}
    F_{\rm s} = \frac{k\left(T_{\rm ab}-T_{\rm s} \right)}{R_{\rm ab}-R_{\rm s}},
\label{eq:surface_heat_flux}
\end{equation}
with $\alpha$ being the volumetric thermal expansion coefficient, $\eta$ the dynamic viscosity, $k$ the thermal conductivity, ${\rm Ra}_{\rm cr}$ the critical Rayleigh number (${\sim} 10^{3}$), and $F_{\rm s}$ the surface heat flux. The viscosity and thermal conductivity are given by Chapman-Enskog theory \citep{Chapman1970}:
\begin{equation}
    k = \frac{1.8}{\gamma-1}\left(\frac{d}{\angstrom}\right)^{-2}\left(\frac{T}{1000}\right)^{\frac{1}{2}}\left(\frac{\mu}{{\rm amu}}\right)^{-\frac{1}{2}}
\label{eq:chapman_conductivity}
\end{equation}
and
\begin{equation}
    \eta = 8.4{\times}10^{-5}\left(\frac{d}{\angstrom}\right)^{-2}\left(\frac{T}{1000}\right)^{\frac{1}{2}}\left(\frac{\mu}{{\rm amu}}\right)^{\frac{1}{2}}.
\end{equation}
Equations~\ref{eq:critical_Rayleigh_number} and \ref{eq:surface_heat_flux} can be solved for the temperature contrast and the boundary layer thickness:
\begin{equation}
    T_{\rm s} = T_{\rm ab}+\left(\frac{{\rm Ra}_{\rm cr} \eta_{\rm ab} F_{\rm s}^{3}}{ \rho_{\rm ab}^{2} c_{\rm p,ab} g_{\rm ab} \alpha_{\rm ab} k_{\rm ab}^{2}} \right)^{\frac{1}{4}}
\label{eq:temperature_through_atmosphere}
\end{equation}
and
\begin{equation}
    R_{\rm ab} = R_{\rm s}+\left(\frac{{\rm Ra}_{\rm cr} \eta_{\rm ab} k_{\rm ab}^{2}}{\rho_{\rm ab}^{2} c_{\rm p,ab} F_{\rm s} g_{\rm ab} \alpha_{\rm ab}}\right)^{\frac{1}{4}}.
\end{equation}

\subsubsection{Optical depth}

Atmospheres have a thermal blanketing effect that retards the cooling rate of the central nucleus \citep{Mizuno1980(2),Matsui1986,Abe1997,Lupu2014}. Conservation of energy requires that the heat flow through the top boundary layer of the magma ocean and the bottom atmospheric boundary layer matches the outward radiation through the atmosphere. In this section, we describe our model for determining the optical depth of the atmosphere, with which thermal blanketing and radiative cooling are calculated during each regime of atmospheric evaporation. We begin with the temperature approximation for an irradiated gray atmosphere with an outward heat flux \citep{Guillot2010}
\begin{equation}
    T^{4} = \frac{3}{4\sigma}F\left(\tau+\frac{2}{3}\right)+\frac{2+\sqrt{3}}{4}T^{4}_{\rm eq},
\label{eq:Guillot_model}
\end{equation}
where $T_{\rm eq}$ is the equilibrium temperature
\begin{equation}
    T_{\rm eq} = T_{\ast}\left(1-A\right)^{\frac{1}{4}}\sqrt{\frac{a}{2R_{\ast}}},
\end{equation}
$T_{\ast}$ and $R_{\ast}$ are the star's effective temperature and radius, respectively, and $A$ is the Bond albedo of the planet \citep[e.g.,][]{Modirrousta2021}. We evaluate equation~\ref{eq:Guillot_model} at the Bondi radius for regime one, whereas in regimes two and three, it is evaluated at the radiative-convective boundary.

For regime one, equation~\ref{eq:Guillot_model} is solved for the outward flux at the Bondi radius,
\begin{equation}
    F_{\rm B}=\frac{4\sigma}{3\tau_{\rm B}+2}\left(T_{\rm B}^{4}-\frac{2+\sqrt{3}}{4}T^{4}_{\rm eq}\right),
\label{eq:Bondi_flux}
\end{equation}
with $\sigma$ being the Stefan-Boltzmann constant, and $\tau_{\rm B}$ the optical depth at the Bondi radius. The parameter $\tau_{\rm B}$ can only be evaluated by considering the equation of state of the escaping winds. Two idealized cases can be conceived: adiabatic and isothermal outflowing gas. The adiabatic solution is given by (see appendix for derivation)
\begin{equation}
    \frac{u^{2}}{u^{2}_{\rm B}}-\ln{\left(\frac{u^{2}}{u^{2}_{\rm B}}\right)} = 4 \ln{\left(\frac{r}{R_{\rm B}}\right)}+1,
\label{eq:adiabatic_steady_state}
\end{equation}
and the isothermal solution is \citep{Parker1964}
\begin{equation}
    \frac{u^{2}}{u^{2}_{\rm B}}-\ln{\left(\frac{u^{2}}{u^{2}_{\rm B}}\right)} = 4 \ln{\left(\frac{r}{R_{\rm B}}\right)}+4\frac{R_{\rm B}}{r}-3.
\end{equation}
\begin{figure}[ht]
    \centering
    \includegraphics[scale = 0.9]{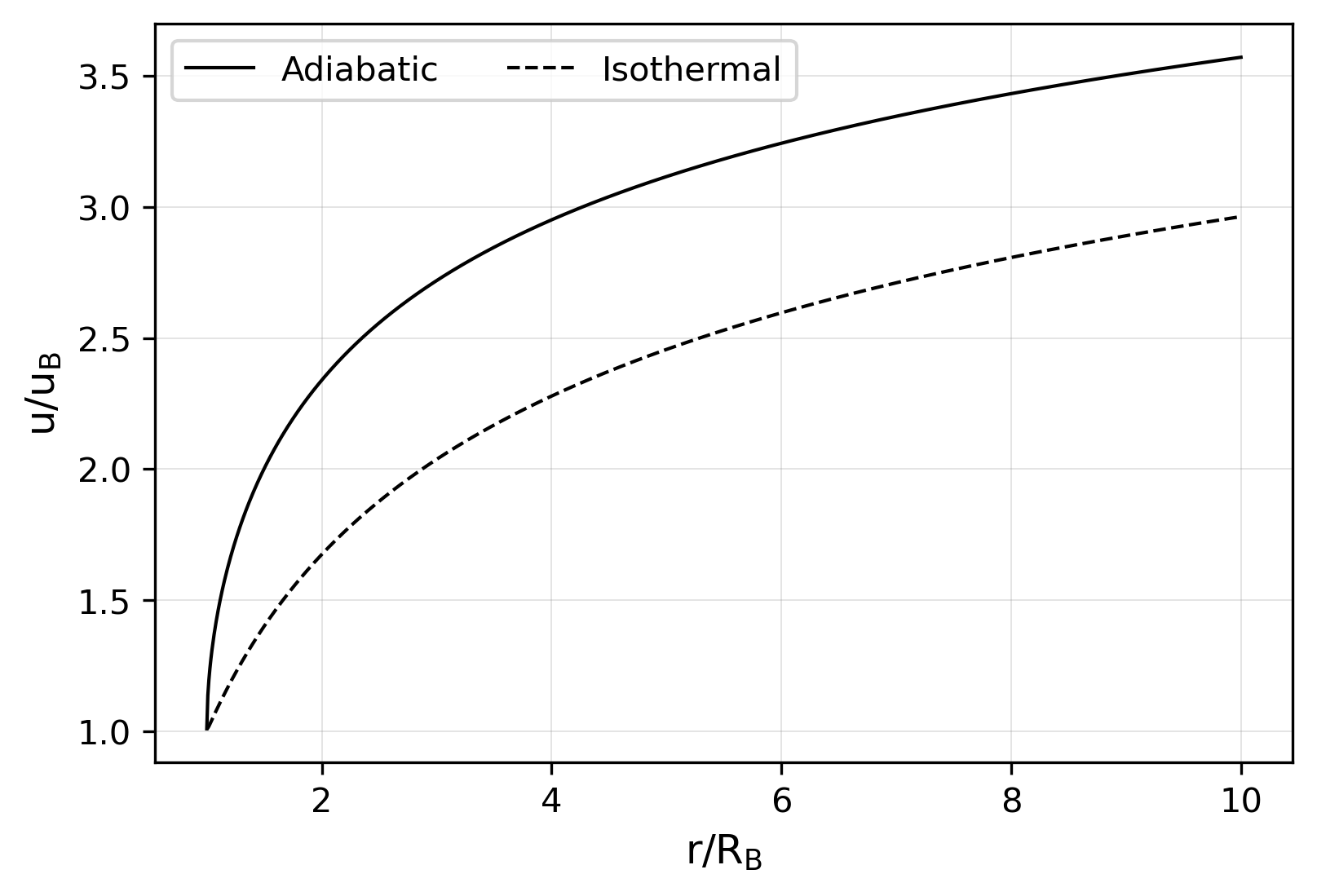}
    \caption{The velocity of outflowing winds for the adiabatic (solid) and isothermal (dashed) cases. The radius and velocity have been divided by their corresponding values at the Bondi radius, respectively.}
    \label{fig:outflow}
\end{figure}

In real systems, the velocity of outflowing gases will be located between the adiabatic and isothermal curves shown in Figure~\ref{fig:outflow}. The limiting behavior of the adiabatic and isothermal cases are $u {\in} \mathcal{O}\left[\ln{\left(r\right)^{1/2}}\right]$. By incorporating this limiting behavior into the equation for mass conservation (i.e., $\rho u r^{2} {=} \rho_{\rm B} u_{\rm B} R_{\rm B}^2$), it can be shown that $\rho{\in}\mathcal{O}\left(r^{-2}\right)$ for both cases. This suggests that the optical depth at the Bondi radius may be approximated as
\begin{equation}
\begin{split}
    \tau_{\rm B} &=\int^{\infty}_{R_{\rm B}} \kappa_{\rm th} \rho dr \\
    &\approx \int^{\infty}_{R_{\rm B}} \kappa_{\rm th} \rho_{\rm B} \left(\frac{R_{\rm B}}{r}\right)^{2} dr\\
    &\approx C_{\rm w} \kappa_{\rm th} \rho_{\rm B} R_{\rm B},
\end{split}
\label{eq:hydrodynamic_case}
\end{equation}
where $C_{\rm w}$ is a correction factor that is found as follows. There are two equations for the optical depth of an atmosphere; one for a hydrodynamic gas and the other for a hydrostatic one. The hydrodynamic approximation is given by equation~\ref{eq:hydrodynamic_case}, and the hydrostatic case has an exact solution given by (see appendix for derivation)
\begin{equation}
    \tau_{\rm B} = \frac{\kappa_{\rm th}\rho_{\rm B}k_{\rm B}T_{\rm B}R_{\rm B}^{2}}{GM_{\rm n}\mu_{\rm B}}.
\label{eq:hydrostatic_case}
\end{equation}
Equations~\ref{eq:hydrodynamic_case} and \ref{eq:hydrostatic_case} apply above and below the Bondi radius respectively. Continuity requires both equations to match at the Bondi radius, so $C_{\rm w} {\equiv} 2/\gamma$. Equations~\ref{eq:hydrodynamic_case} and \ref{eq:Bondi_flux} therefore become
\begin{equation}
    \tau_{\rm B} \approx \frac{2}{\gamma}\kappa_{\rm th} \rho_{\rm B} R_{\rm B}
\label{eq:optical_depth_Bondi}
\end{equation}
and
\begin{equation}
    F_{\rm B}\approx \frac{2\gamma\sigma}{3\kappa_{\rm th} \rho_{\rm B} R_{\rm B}+\gamma}\left(T_{\rm B}^{4}-\frac{2+\sqrt{3}}{4}T^{4}_{\rm eq}\right),
\label{eq:F_Bondi}
\end{equation}
respectively. 

For regimes two and three, equation~\ref{eq:Guillot_model} is solved for the outward heat flux at the radiative-convective boundary,
\begin{equation}
    F_{\rm rcb}=\frac{4\sigma}{3\tau_{\rm rcb}+2}\left(T_{\rm rcb}^{4}-\frac{2+\sqrt{3}}{4}T^{4}_{\rm eq}\right).
\label{eq:F_rcb}
\end{equation}
The radiative-convective boundary is the location where the outward heat flux is equal to the incoming radiant flux from the star, so from equation~\ref{eq:Guillot_model} one gets,
\begin{equation}
    \frac{3}{4\sigma}F_{\rm rcb}\left(\tau_{\rm rcb}+\frac{2}{3}\right) = \frac{2+\sqrt{3}}{4}T^{4}_{\rm eq}.
\label{eq:equal_flux}
\end{equation}
The temperature at the radiative-convective boundary is found by combining equations~\ref{eq:equal_flux} and \ref{eq:Guillot_model},
\begin{equation}
    T_{\rm rcb} = \left(\frac{2+\sqrt{3}}{2}\right)^{\frac{1}{4}} T_{\rm eq},
\label{eq:Trcb_value}
\end{equation}
which can be inserted into equation~\ref{eq:F_rcb},
\begin{equation}
    F_{\rm rcb}=\frac{2+\sqrt{3}}{3\tau_{\rm rcb}+2} \sigma T^{4}_{\rm eq}.
\label{eq:F_rcb_complete}
\end{equation}
The optical depth at the radiative-convective boundary, $\tau_{\rm rcb}$, is found by integrating $d\tau {=} \rho(r) \kappa dr$ across the atmosphere and escaping winds.

\subsection{Opacity treatment}

We adopt a grey atmospheric model with constant thermal and XUV opacities. Gas opacities are temperature, density, and wavelength dependent, with chemistry \citep{Freedman2014}, ionization \citep{Chadney2022}, and free-floating grains \citep{Henning1996} further increasing their uncertainty. XUV opacities are relevant only when internal temperatures are low enough for the planet to experience XUV-induced photoevaporation (section~\ref{sec:regime_3}), so we adopt a constant value of $\rm 10^{5}~m^{2}~kg^{-1}$ \citep[converted from the cross-sectional area data of molecular hydrogen in Figure~7 of][]{Chadney2022} because thermal effects are probably minor. Thermal opacities are, however, expected to vary through the evolution of the planet. Immediately after the last giant impact, surface temperatures are high enough for bound-free, free-free, and Thompson scattering opacities to apply \citep{Hayashi1962,Cox1976}. At lower temperatures, the opacity is dominated by hydrogen anions \citep{Wishart1979} and grains \citep{Henning1996}. We choose an average value of $\rm 1~m^{2}~kg^{-1}$, which is intermediate between the very high opacities arising from the high initial temperatures and densities \citep[$\rm {\sim}10^{3}~m^{2}~kg^{-1}$;][]{Rogers1996,Iglesias1996} and the lower opacities when temperatures and densities are lower \citep[$\rm {\sim} 10^{-3}~m^{2}~kg^{-1}$;][]{Henning1996,Freedman2014}. Adopting different values for the opacity does not invalidate the findings of this study, that is, there are three regimes of atmospheric evaporation for super-Earths and sub-Neptunes. The existence of regime one is independent of the opacity of the atmosphere, as demonstrated in our model description in section~\ref{sec:regime_1}. The formation of regime two (equation~\ref{eq:first_criterion}) is also independent of the opacity. Regime three is the only regime whose existence depends on both opacities; the thermal opacity sets the location of the photosphere and the XUV opacity determines how many ionizing photons are absorbed in the thermosphere. Therefore, the choice of opacity will only alter the model details, such as the temperature and pressure profiles and the effects of thermal blanketing, but not the main findings.

\subsection{Cloud formation}
\label{sec:cloud_formation}

The high internal temperatures immediately after a giant impact will inhibit condensation. Whereas clouds can form from any condensable species, $\rm SiO$ and $\rm SiO_{2}$ are the most probable candidates for hot exoplanets \citep{Schaefer2012,Ito2015}. Water clouds are not possible because condensation cannot occur above its critical point temperature of $647~{\rm K}$, which is significantly lower than the temperatures of interest to this study. The thermodynamic data of \citet{Chase1985} indicate that thermal decomposition is favored for an Si-O gas mixture above $\rm 6000~{\rm K}$, with the critical point temperature being approximately $\rm 6300~K$ \citep{Iosilevskiy2013,Connolly2016,Xiao2018}. Condensation is therefore not possible above $\rm 6000~K$. At lower temperatures, silicate clouds will form only if the vapor pressure is lower than the partial pressure:
\begin{equation}
    P_{\rm v}(T) \leq fP,
\label{eq:condition_cloud}
\end{equation}
where $P_{\rm v}(T)$ and $f$ are the vapor pressure and mole fraction of the species of interest respectively. As previously stated, we do not consider atmospheric enrichment mechanisms in this paper. The parameter $f$ is therefore set by the composition of the nebular gas that formed the primordial atmosphere, which is negligible and of the order $10^{-5}$ for silicon \citep{Lodders2003,Lodders2010}. The inclusion of atmospheric enrichment mechanisms will be left for future studies, so we do not include cloud formation in this paper.

\section{Regime one: Fully convecting atmosphere}
\label{sec:regime_1}

In this regime, we model the atmospheric evaporation rates in the immediate aftermath of the last giant impact when the mechanical atmospheric waves have dissipated. The extreme temperatures of regime one allow for a fully convecting atmosphere because the Bondi radius lies within the radiative-convective boundary. Because the photosphere is always located at the radiative-convective boundary or above (in optically thin regions, radiation is more efficient than convection), the atmosphere must be optically thick. Our model for regime one is structured as follows. In section~\ref{sec:regime_one_requirements}, we explain the requirements for regime one to occur. In section~\ref{sec:regime_one_atmospheric_model}, we describe our atmospheric model. In section~\ref{sec:analytic_solution}, we show that radiative cooling is negligible during regime one, so mass loss will be the major mechanism that modifies the temperature profile of the planet.

\subsection{Regime one requirements}
\label{sec:regime_one_requirements}

An atmosphere must satisfy a minimum mass requirement for it to be optically thick at the Bondi radius and therefore to be allowed to enter regime one where it is fully convecting. We begin by considering a highly luminous planet with an atmosphere that is marginally below the minimum mass requirement. The atmosphere will therefore be fully radiative, with its density distribution below the Bondi radius approximately following the barometric formula \citep{Lente2020}. Evaluating the barometric formula at the Bondi radius gives
\begin{equation}
    \rho_{\rm B} = \rho_{\rm s} \exp{\left[-\frac{GM_{\rm n}\bar{\mu}}{k_{\rm B}T_{\rm B}}\left(\frac{1}{R_{\rm s}}-\frac{1}{R_{\rm B}}\right)\right]},
\label{density_appendix_1}
\end{equation}
where $\rho_{\rm s}$ is the density of gas at the base of the atmosphere. Above the Bondi radius, the density distribution is given by the steady state solution of an adiabatic wind (equation~\ref{eq:adiabatic_steady_state} and conservation of mass). At the Bondi radius, the adiabatic solution gives (see equation~\ref{eq:tau_B}),
\begin{equation}
    \rho_{\rm B} = \frac{\gamma}{3\kappa_{\rm th}R_{\rm B}}.
\label{density_appendix_2}
\end{equation}
Combining equations~\ref{density_appendix_1} and \ref{density_appendix_2},
\begin{equation}
    M_{\rm a,min} = \frac{8 \pi R_{\rm s}^{4}}{3\kappa_{\rm th}R_{\rm B}^{2}}\exp{\left[\frac{\gamma}{2}\left(\frac{R_{\rm B}}{R_{\rm s}}-1\right) \right]}.
\end{equation}

\begin{figure}[htbp]
    \centering
    \includegraphics{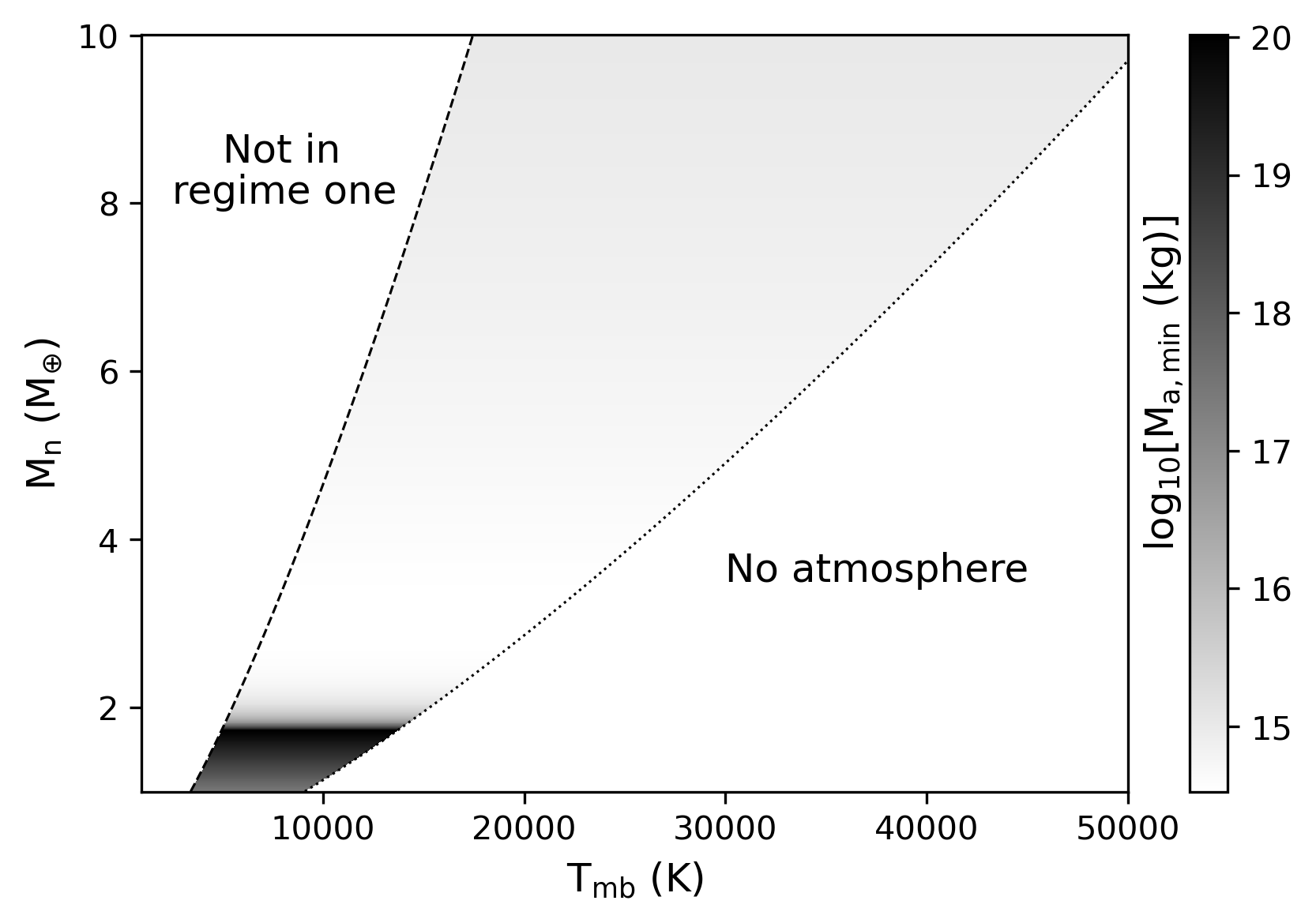}
    \caption{The minimum mass required for an atmosphere to be optically thick at the Bondi radius and therefore be allowed to convect throughout, as a function of the mass of the nucleus and the surface temperature. The dashed and dotted lines are for the lower and upper temperature limits shown in equation~\ref{eq:first_criterion}, respectively.}
    \label{fig:min_atm}
\end{figure}

In addition to the minimum mass requirement, there is also a temperature requirement. Total atmospheric convection (from $R_{\rm ab}$ to $R_{\rm B}$) occurs only when convection is the most efficient energy transfer mechanism in the atmosphere. The transition temperature (i.e., from convection to radiation) is the location where the radiative heat component is equal to the outward heat flux, as shown in equation~\ref{eq:equal_flux}. The transition temperature is therefore 
\begin{equation}
    T_{\rm B} = \left(\frac{2+\sqrt{3}}{2}\right)^{\frac{1}{4}} T_{\rm eq}.
\label{eq:transition_temperature}
\end{equation}
The temperature at the bottom of the convecting atmosphere is found by following the adiabat (equation~\ref{eq:Temperature_profile}) to the radius $R_{\rm ab}$. Combining this with the maximum temperature at which an atmosphere remains gravitationally bound, we obtain the temperature range relevant to regime one:
\begin{equation}
    \frac{3-\gamma}{2}\left(\frac{2+\sqrt{3}}{2}\right)^{\frac{1}{4}}T_{\rm eq} + \frac{\gamma-1}{\gamma}\frac{GM_{\rm n}\bar{\mu}}{k_{\rm B}R_{\rm ab}} < T_{\rm ab} < \frac{2}{\gamma} \frac{GM_{\rm n}\bar{\mu}}{k_{\rm B}R_{\rm ab}}.
\label{eq:first_criterion}
\end{equation}
The right-hand side is found by rearranging equation~\ref{eq:Bondi_radius} for the temperature and setting the Bondi radius equal to the radius $R_{\rm ab}$.

\subsection{Atmospheric Model}
\label{sec:regime_one_atmospheric_model}

\begin{figure}[ht]
    \centering
    \includegraphics[scale = 0.3]{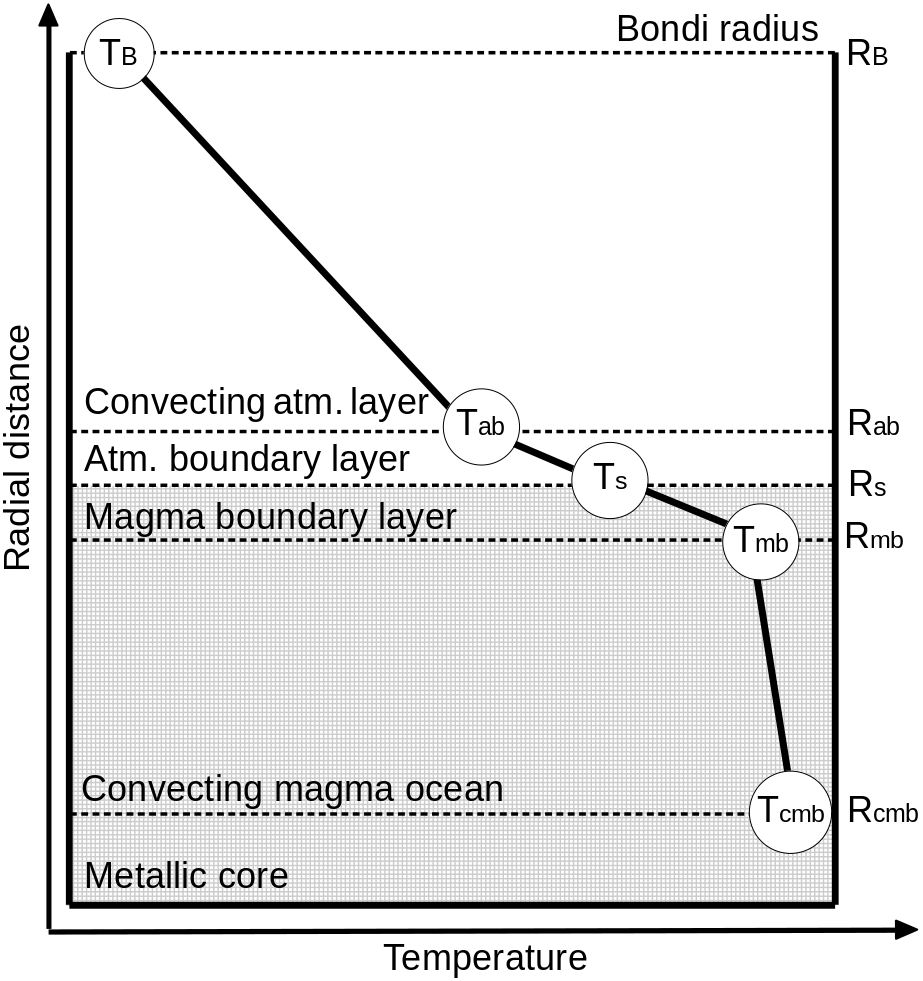}
    \caption{The expected temperature profile of the planet during regime one. Temperature and distance are not to scale; $T_{\rm B}$, $T_{\rm ab}$, $T_{\rm s}$ $T_{\rm mb}$, and $T_{\rm cmb}$ are the temperatures at the Bondi radius, top of the atmospheric boundary layer, surface of the nucleus, top of the convecting magma ocean, and metallic core-magma ocean boundary layer, respectively. $R$ with the relevant subscript marks the radius at which each temperature is defined.}
    \label{fig:regime_one}
\end{figure}

We begin with the equation for hydrostatic equilibrium
\begin{equation}
    \frac{dP}{dr} = -\rho g,
\end{equation}
where $g$ is the gravitational acceleration. For a convecting unsaturated (i.e., cloudless) ideal gas, $\rho$ and $P$ are related by the previously mentioned isentropic relations (equation~\ref{eq:isentropic_relations}), which can be combined with $\rho_{\rm B} {=} P_{\rm B} \bar{\mu}/(k_{\rm B}T_{\rm B})$ to get
\begin{equation}
    \frac{dP}{dr} = -\frac{P_{\rm B}\bar{\mu}}{k_{\rm B}T_{\rm B}} \left(\frac{P}{P_{\rm B}} \right)^{\frac{1}{\gamma}} \frac{GM_{\rm n}}{r^{2}}.
\label{eq:initial_density_equation}
\end{equation}
We assume that the atmospheric mass is negligible compared to the total planetary mass, and we use the bulk mean molecular mass $\bar{\mu}$ because the entire atmosphere is well-mixed in this regime. We integrate both sides and solve for $P/P_{\rm B}$:
\begin{equation}
    \left(\frac{P}{P_{\rm B}}\right)^{\frac{\gamma-1}{\gamma}} = 1+\frac{\gamma-1}{\gamma}\frac{GM_{\rm n}\bar{\mu}}{k_{\rm B}T_{\rm B}} \left(\frac{1}{r} -\frac{1}{R_{\rm B}}\right).
\label{eq:pressure_ratio}
\end{equation}
Using the relations of an isentropic gas, we may also derive the temperature profile of a fully adiabatic atmosphere:
\begin{equation}
    T = T_{\rm B}\left[1+\frac{\gamma-1}{\gamma}\frac{GM_{\rm n}\bar{\mu}}{k_{\rm B}T_{\rm B}} \left(\frac{1}{r} -\frac{1}{R_{\rm B}}\right)\right].
\label{eq:Temperature_profile}
\end{equation}
By setting $T = T_{\rm ab}$ and $r=R_{\rm ab}$, the Bondi temperature is found:
\begin{equation}
  T_{\rm B} = \frac{2}{3-\gamma} \left(T_{\rm ab}-\frac{\gamma-1}{\gamma}\frac{GM_{\rm n}\bar{\mu}}{k_{\rm B}R_{\rm ab}}\right).
\label{eq:T_B}
\end{equation}
Using the isentropic gas relations with equation~\ref{eq:pressure_ratio}, the density profile is found as
\begin{equation}
    \rho = \rho_{\rm B}\left(\frac{3-\gamma}{2} + \frac{\gamma-1}{2}\frac{R_{\rm B}}{r} \right)^{\frac{1}{\gamma-1}}.
\label{eq:density_profile}
\end{equation}
The atmospheric mass is therefore
\begin{equation}
\begin{split}
    M_{\rm a} &= 4 \pi \int^{R_{\rm B}}_{R_{\rm ab}} \rho r^{2} dr \\
    &= 4 \pi \rho_{\rm B} \int^{R_{\rm B}}_{R_{\rm ab}} \left(\frac{3-\gamma}{2} + \frac{\gamma-1}{2}\frac{R_{\rm B}}{r} \right)^{\frac{1}{\gamma-1}} r^{2} dr.
\end{split}
\label{eq:atmospheric_mass}
\end{equation}
Rearranging equation~\ref{eq:atmospheric_mass} for $\rho_{\rm B}$ gives the Bondi density:
\begin{equation}
    \rho_{\rm B} = M_{\rm a}\left[4 \pi \int^{R_{\rm B}}_{R_{\rm ab}} \left(\frac{3-\gamma}{2} + \frac{\gamma-1}{2}\frac{R_{\rm B}}{r} \right)^{\frac{1}{\gamma-1}} r^{2} dr\right]^{-1}.
\label{eq:Bondi_density}
\end{equation}
Finally, the mass loss rate at the Bondi radius may be calculated as
\begin{equation}
    \dot{M}_{\rm B}= 4 \pi \xi_{1} R^{2}_{\rm B} \rho_{\rm B} u_{\rm B},
\label{eq:Bondi_limited}
\end{equation}
where $u_{\rm B}$ is the speed of sound, 
\begin{equation}
    u_{B} = \sqrt{\frac{\gamma k_{B}T_{B}}{\bar{\mu}}},
\label{eq:velocity}
\end{equation}
and $\xi_{1}$ is the efficiency of mass loss, which is one half because half of all particles are scattered radially outward.

A planet will lose energy through mass loss (i.e., decreasing the gravitational potential energy) and radiative cooling. The balance of these two effects will play an critical role in the evolution of the interior temperature profile of a planet and in the amount of mass loss taking place. As shown in the following section, this balance is found by comparing the equation for radiative cooling in regime one (equation~\ref{eq:F_Bondi}) with the energy loss from atmospheric outflow.

\subsection{Critical atmospheric mass for regime one}
\label{sec:analytic_solution}

\begin{figure}[htbp]
    \centering
    \includegraphics[scale = 0.9]{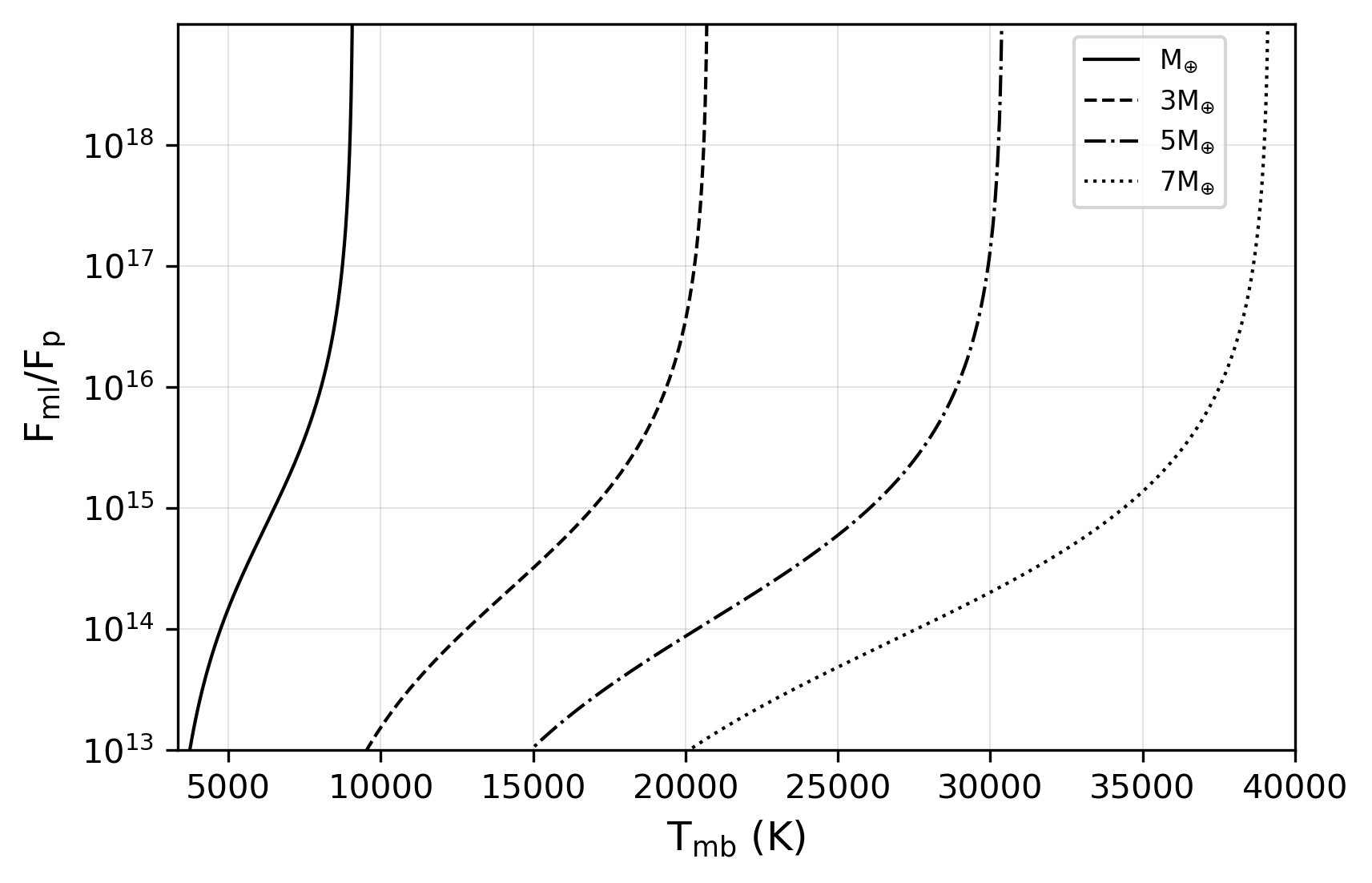}
    \caption{The ratio of the mass loss and radiative cooling heat fluxes as a function of the magma ocean temperature for different masses of the nucleus. For each case, the atmospheric mass is $1\%$ of the total planetary mass.}
    \label{fig:ratio}
\end{figure}

To determine the importance of radiative cooling in influencing the evolution of a planet during regime one, we compare equation~\ref{eq:F_Bondi} with the loss of energy from atmospheric evaporation (see appendix for derivation),
\begin{equation}
    F_{\rm ml} \approx \frac{GM_{\rm n}\dot{M}_{\rm a}}{4\pi R^{3}_{\rm B}}.
\end{equation}
Figure~\ref{fig:ratio} shows the ratio of the mass loss and radiative cooling energy fluxes for different parameter choices. The mass loss energy flux is always significantly larger than that of radiative cooling.

\begin{figure}[ht]
    \centering
    \includegraphics[scale = 0.9]{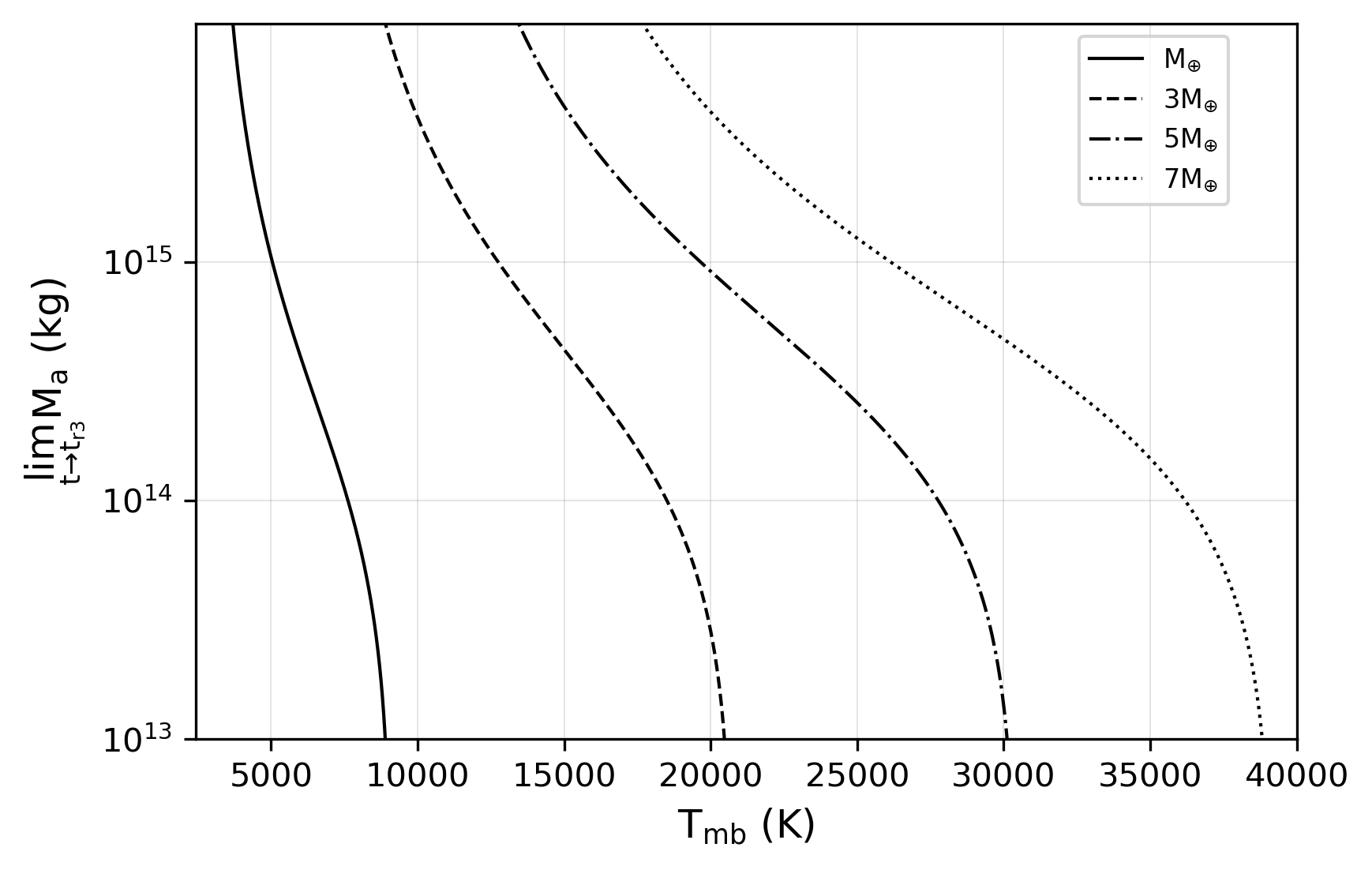}
    \caption{The atmospheric mass as a function of the magma ocean temperature and the mass of the nucleus at which the Bondi radius becomes optically thin and energy can be efficiently lost through radiation.}
    \label{fig:crossover_mass}
\end{figure}

As shown in Figure~\ref{fig:ratio}, radiative cooling is small compared to the effects of mass loss. If radiative cooling is taken to be negligible, the atmosphere will remain in a fully convective state until the density at the Bondi radius is low enough for the optical depth to equal two thirds:
\begin{equation}
    \tau_{\rm B} \approx \frac{2}{\gamma}\kappa_{\rm th} \rho_{\rm B} R_{\rm B} = \frac{2}{3}.
\label{eq:tau_B}
\end{equation}
Equation~\ref{eq:tau_B} can be combined with equation~\ref{eq:Bondi_density} to find the critical mass at which the atmosphere will form a radiative region,
\begin{equation}
    \lim_{\mathcal{R}_{1}\to \mathcal{R}_{3}}{M_{\rm a}} = \frac{4 \pi \gamma}{3\kappa_{\rm th} R_{\rm B}} \int^{R_{\rm B}}_{R_{\rm ab}} \left(\frac{3-\gamma}{2} + \frac{\gamma-1}{2}\frac{R_{\rm B}}{r} \right)^{\frac{1}{\gamma-1}} r^{2} dr,
\end{equation}
where $\mathcal{R}_{1}$ and $\mathcal{R}_{3}$ are regimes one and three respectively. Regime two will not be traversed because it requires the planet to have lower internal temperatures while the atmosphere remains optically thick. This is not possible because, as evidenced by Figure~\ref{fig:ratio}, mass loss is substantially greater than cooling so that the atmosphere becomes optically thin before the interior has had enough time to cool. Figure~\ref{fig:crossover_mass} shows the final atmospheric mass of an an exoplanet in regime one with initial temperature $T_{\rm mb}$. After reaching this critical mass, the atmosphere will form a radiative layer with a photosphere.

Though it has been shown that radiative cooling is negligible in regime one, it is still incorporated in our simulations. Our approach is to first set the temperature of the magma ocean, with which the temperatures across the top magma ocean boundary layer (equation~\ref{eq:flux_through_magma}), bottom atmospheric boundary layer (equation~\ref{eq:temperature_through_atmosphere}), and the temperature at the Bondi radius (equation~\ref{eq:T_B}) are found through iteration to attain self-consistency. Through the Bondi radius temperature, the outward heat flux is calculated with equation~\ref{eq:F_Bondi}, which is then used to determine the temperature decrease of the magma ocean with equation~\ref{eq:interior_cooling}. The process is then repeated to track the thermal evolution of the magma ocean.

\section{Regime two: Formation of a radiative layer}
\label{sec:regime_2}

In the following, we describe the physics of regime two. In sections~\ref{sec:regime_two_requirements} and \ref{sec:XUV_importance}, we explain why regime two is intermediate between regime one and regime three, that is, why it is cold enough for a radiative region to form (i.e., the radiative-convective boundary lies within the Bondi radius) but too hot for a photosphere (where $\tau_{\rm th}{=}2/3$). The atmosphere will no longer follow an adiabatic temperature profile through to the Bondi radius because a radiative region forms above the radiative-convective boundary, where the temperature is set by radiative equilibrium. In section~\ref{sec:regime_2_atmospheric_model}, we describe our strategy for modeling the temperature and density profiles of the atmosphere to find the conditions at the Bondi radius, with which the mass loss rate can be found.

\subsection{Regime two requirements}
\label{sec:regime_two_requirements}

A radiative region forms in regime two, so internal temperatures are below those given in equation~\ref{eq:first_criterion}. We therefore focus on temperatures of
\begin{equation}
    T_{\rm ab} < \frac{3-\gamma}{2}\left(\frac{2+\sqrt{3}}{2}\right)^{\frac{1}{4}}T_{\rm eq} + \frac{\gamma-1}{\gamma}\frac{GM_{\rm n}\bar{\mu}}{k_{\rm B}R_{\rm ab}}.
\end{equation}
Because regime two is still too hot for a photosphere to form, the density of the Bondi radius must be higher than that of the photosphere. This latter condition can be expressed in terms of a minimum Bondi temperature that is found by solving equation~\ref{eq:optical_depth_Bondi} for an optical depth of ${2/3}$,
\begin{equation}
    T_{\rm B} > \frac{6 \kappa_{\rm th} G M_{\rm n}\bar{\mu}\rho_{\rm B}}{\gamma^{2}k_{\rm B}}.
\label{eq:Bondi_temperature_condition}
\end{equation}
The corresponding basal atmospheric temperature at which equation~\ref{eq:Bondi_temperature_condition} occurs depends on the mass of the nucleus and atmosphere and on the thermodynamic conditions of the interior.

\subsection{The importance of XUV irradiation in regime two}
\label{sec:XUV_importance}

XUV driven mass loss can occur only above the photosphere; below it, the atmosphere is optically thick to thermal photons, so the Stefan-Boltzmann law holds. In other words, any XUV photons absorbed below the photosphere will cause the local gas to heat up and radiate the excess energy like a blackbody. The inefficiency of XUV-induced heating in optically thick regions can be demonstrated through the following example. Consider a planet orbiting a star that has zero brightness in its XUV bands. The effective flux of the planet would be given by the Stefan-Boltzmann law
\begin{equation}
    F_{\rm eff} = \sigma T_{\rm eff}^{4}.
\end{equation}
Adding the XUV component results in a temperature increase
\begin{equation}
    F_{\rm eff}+F_{\rm XUV} = \sigma \left(T_{\rm eff} + \Delta T\right)^{4},
\end{equation}
whence
\begin{equation}
    \frac{\Delta T}{T_{\rm eff}} = \left(1+\frac{F_{\rm XUV}}{F_{\rm eff}}\right)^{\frac{1}{4}}-1.
\end{equation}
The maximum temperature change will occur when the planet has no internal energy, so the effective flux is caused entirely by stellar irradiation. In this limiting case, the ratio of the XUV and effective fluxes is defined by the luminosity of the host star at different wavelengths,
\begin{equation}
    \frac{\Delta T}{T_{\rm eff}} = \left(1+\frac{L_{\rm XUV}}{L_{\rm bol}-L_{\rm XUV}}\right)^{\frac{1}{4}}-1,
\end{equation}
where $L_{\rm bol}$ is the bolometric luminosity (the total luminosity of the star not including the emission of neutrinos). Even when a star is very young and brightest in its XUV bands, the luminosity ratio is at most ${\sim} 10^{-3}$ \citep{Vilhu1987,Penz2008(1),Penz2008(2)}. The maximum value of $\Delta T/T_{\rm eff}$ is therefore of the order $10^{-4}$, so the influence of XUV irradiation on the temperature can be ignored below the photosphere. We therefore do not incorporate the effects of XUV irradition in this regime. 

\subsection{Atmospheric model}
\label{sec:regime_2_atmospheric_model}

\begin{figure}[ht]
    \centering
    \includegraphics[scale = 0.3]{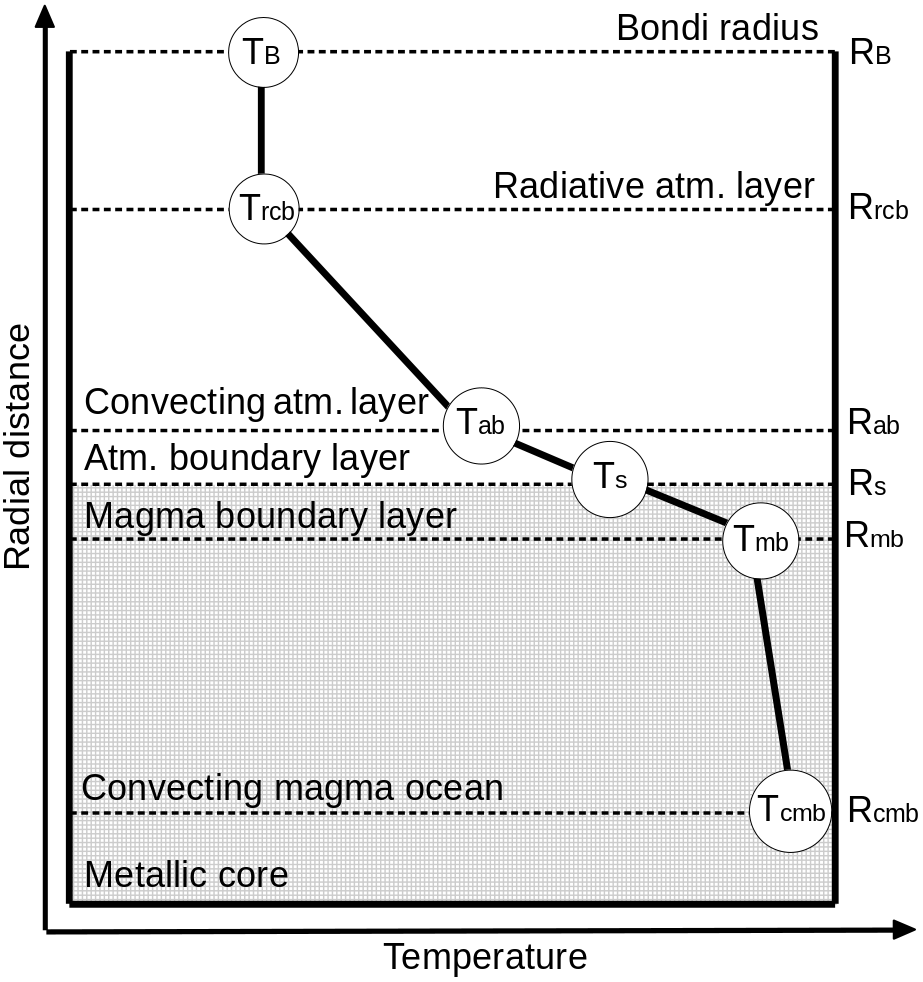}
    \caption{The expected temperature profile of the planet during regime two. Temperature and distance are not to scale; $T_{\rm B}$, $T_{\rm rcb}$, $T_{\rm ab}$, $T_{\rm s}$ $T_{\rm mb}$, and $T_{\rm cmb}$ are the temperatures at the Bondi radius, radiative-convective boundary, top of the atmospheric boundary layer, surface of the nucleus, top of the convecting magma ocean, and metallic core-magma ocean boundary layer, respectively. $R$ with the relevant subscript marks the radius at which each temperature is defined.}
    \label{fig:regime_two}
\end{figure}

In a similar manner to our approach in section~\ref{sec:regime_1}, we will find the density and temperature at the Bondi radius to estimate the mass loss rates. In this regime, a radiative region will form from the radiative-convective boundary $R_{\rm rcb}$ to the Bondi radius $R_{\rm B}$. The radiative-convective boundary is located where the outward heat flux is equal to the radiant flux (see equation~\ref{eq:transition_temperature})
\begin{equation}
    T_{\rm rcb} = \left(\frac{2+\sqrt{3}}{2}\right)^{\frac{1}{4}} T_{\rm eq}.
\label{eq:T_rcb_T_eq}
\end{equation}
The density at the radiative-convective boundary is found in a similar manner to equation~\ref{eq:density_profile}, but by changing the boundary conditions from those corresponding at the Bondi radius to those at the radiative-convective boundary:
\begin{equation}
    \rho = \rho_{\rm rcb}\left[1+\frac{\gamma-1}{\gamma}\frac{GM_{\rm n}\bar{\mu}}{k_{\rm B}T_{\rm rcb}} \left(\frac{1}{r}-\frac{1}{R_{\rm rcb}} \right)\right]^{\frac{1}{\gamma-1}}.
\label{eq:rho_rcb}
\end{equation}
The temperature profile is found by adapting equation~\ref{eq:Temperature_profile}
\begin{equation}
    T = T_{\rm rcb}\left[1+\frac{\gamma-1}{\gamma}\frac{GM_{\rm n}\bar{\mu}}{k_{\rm B}T_{\rm rcb}} \left(\frac{1}{r} -\frac{1}{R_{\rm rcb}}\right)\right],
\label{eq:T_rcb}
\end{equation}
which can then be solved for the radius of the radiative-convective boundary
\begin{equation}
    R_{\rm rcb} = R_{\rm ab}\left[1-\frac{\gamma}{\gamma-1}\frac{k_{\rm B}R_{\rm ab}}{GM_{\rm n}\bar{\mu}}\left(T_{\rm ab}-T_{\rm rcb}\right) \right]^{-1}.
\label{eq:R_rcb}
\end{equation}
Combining equations~\ref{eq:T_rcb_T_eq}, \ref{eq:rho_rcb}, and \ref{eq:R_rcb} removes the dependency on $R_{\rm rcb}$ and $T_{\rm rcb}$,
\begin{equation}
    \rho = \rho_{\rm rcb} \left(\frac{2}{2+\sqrt{3}}\right)^{\frac{1}{4\left(\gamma-1\right)}} \left[\frac{T_{\rm ab}}{T_{\rm eq}}-\frac{\gamma-1}{\gamma}\frac{GM_{\rm n}\bar{\mu}}{k_{\rm B} T_{\rm eq}}\left(\frac{1}{R_{\rm ab}}-\frac{1}{r}\right) \right]^{\frac{1}{\gamma-1}}.
\label{eq:convective_density}
\end{equation}
Equation~\ref{eq:convective_density} gives the density profile in the convective section of the atmosphere (from $R_{\rm ab}$ to $R_{\rm rcb}$), so it is not applicable to the the radiative section (from $R_{\rm rcb}$ to $R_{\rm B}$). Above the radiative-convective boundary and below the photosphere, the atmosphere is approximately isothermal because any major temperature anomaly will be radiated away through blackbody radiation. The greatest temperature difference is found by evaluating equation~\ref{eq:Guillot_model} at its extrema. The smallest value is when $F{=}0$, and the largest value occurs when the radiative heat component equals the outward heat flux (equation~\ref{eq:Trcb_value}). The difference between these extrema is less than $20\%$. Because we focus on hot planets with large outward heat fluxes, thermal deviations will be significantly smaller than the theoretical maximum of $20\%$. Under the isothermal assumption, the density profile is given by the barometric formula \citep{Lente2020}, which is derived in the following. We start with the equation for hydrostatic equilibrium, and we replace the density with the pressure by using the ideal gas equation
\begin{equation}
\begin{split}
    \frac{dP}{dr} = -\frac{P \bar{\mu}}{k_{\rm B}T} \frac{GM_{\rm n}}{r^{2}},
\end{split}
\end{equation}
which can be integrated and solved for the density,
\begin{equation}
    \rho = \rho_{\rm rcb} \exp{\left[-\frac{GM_{\rm n}\bar{\mu}}{k_{\rm B}T_{\rm rcb}}\left(\frac{1}{R_{\rm rcb}}-\frac{1}{r}\right) \right]}.
\label{eq:barometric_formula}
\end{equation}
The dependency on $T_{\rm rcb}$ and $R_{\rm rcb}$ can be removed by substituting in equation~\ref{eq:T_rcb_T_eq} and \ref{eq:R_rcb}, leading to
\begin{equation}
    \rho = \rho_{\rm rcb} \exp{\left[\frac{\gamma}{\gamma-1}\frac{T_{\rm ab}-\left(\frac{2+\sqrt{3}}{2}\right)^{\frac{1}{4}}T_{\rm eq}}{\left(\frac{2+\sqrt{3}}{2}\right)^{\frac{1}{4}}T_{\rm eq}}-\left(\frac{2}{2+\sqrt{3}}\right)^{\frac{1}{4}}\frac{GM_{\rm n}\bar{\mu}}{k_{\rm B}T_{\rm eq}}\left(\frac{1}{R_{\rm ab}}-\frac{1}{r}\right) \right]}.
\label{eq:radiative_density}
\end{equation}
Equation~\ref{eq:convective_density} and \ref{eq:radiative_density} apply when $\rho \geq \rho_{\rm rcb}$ and $\rho < \rho_{\rm rcb}$, respectively, so the governing equation for the density is given by
\begin{equation}
    \rho = \begin{cases}
  \rho_{\rm rcb} \left(\frac{2}{2+\sqrt{3}}\right)^{\frac{1}{4\left(\gamma-1\right)}} \left[\frac{T_{\rm ab}}{T_{\rm eq}}-\frac{\gamma-1}{\gamma}\frac{GM_{\rm n}\bar{\mu}}{k_{\rm B} T_{\rm eq}}\left(\frac{1}{R_{\rm ab}}-\frac{1}{r}\right) \right]^{\frac{1}{\gamma-1}} & r \leq R_{\rm rcb} \\
  \rho_{\rm rcb} \exp{\left[\frac{\gamma}{\gamma-1}\frac{T_{\rm ab}-\left(\frac{2+\sqrt{3}}{2}\right)^{\frac{1}{4}}T_{\rm eq}}{\left(\frac{2+\sqrt{3}}{2}\right)^{\frac{1}{4}}T_{\rm eq}}-\left(\frac{2}{2+\sqrt{3}}\right)^{\frac{1}{4}}\frac{GM_{\rm n}\bar{\mu}}{k_{\rm B}T_{\rm eq}}\left(\frac{1}{R_{\rm ab}}-\frac{1}{r}\right) \right]} & r > R_{\rm rcb}.
\end{cases}
\label{eq:rho_regime_two}
\end{equation}
The density at the radiative-convective boundary is found by rearranging
\begin{equation}
    M_{\rm a} = \int^{R_{\rm B}}_{\rm R_{\rm ab}} 4 \pi r^{2} \rho dr
\end{equation}
so that
\begin{equation}
\begin{split}
    \rho_{\rm rcb} &= \frac{M_{\rm a}}{4 \pi} \left\{\int^{R_{\rm rcb}}_{R_{\rm ab}} r^{2} \left(\frac{2}{2+\sqrt{3}}\right)^{\frac{1}{4\left(\gamma-1\right)}} \left[\frac{T_{\rm ab}}{T_{\rm eq}}-\frac{\gamma-1}{\gamma}\frac{GM_{\rm n}\bar{\mu}}{k_{\rm B} T_{\rm eq}}\left(\frac{1}{R_{\rm ab}}-\frac{1}{r}\right) \right]^{\frac{1}{\gamma-1}}~dr \right. \\ 
    & +\left.\int^{R_{\rm B}}_{R_{\rm rcb}} r^{2} \exp{\left[\frac{\gamma}{\gamma-1}\frac{T_{\rm ab}-\left(\frac{2+\sqrt{3}}{2}\right)^{\frac{1}{4}}T_{\rm eq}}{\left(\frac{2+\sqrt{3}}{2}\right)^{\frac{1}{4}}T_{\rm eq}}-\left(\frac{2}{2+\sqrt{3}}\right)^{\frac{1}{4}}\frac{GM_{\rm n}\bar{\mu}}{k_{\rm B}T_{\rm eq}}\left(\frac{1}{R_{\rm ab}}-\frac{1}{r}\right) \right]}~dr \right\}^{-1}.
\end{split}
\label{eq:rho_rcb_regime_2}
\end{equation}
The density at the Bondi radius can then be found with $\rho_{\rm rcb}$ and equation~\ref{eq:radiative_density}, allowing for an estimation of the mass loss rate using equation~\ref{eq:Bondi_limited}.

To track the thermal evolution of the planet, we adopt the same strategy as with regime one, except that we solve for the conditions at the radiative-convective boundary instead of the Bondi radius. Equation~\ref{eq:F_rcb_complete} provides the outward heat flux, with which planetary cooling is determined using equation~\ref{eq:interior_cooling}.

\section{Regime three: XUV-induced photoevaporation}
\label{sec:regime_3}

In regime three, the atmosphere is cold enough for the radiative-convective boundary and photosphere to form below the Bondi radius. The atmosphere is thus convecting from its base to the radiative-convective boundary, radiative from the radiative-convective boundary to the photosphere, and conducting from the photosphere to the Bondi radius (i.e., the thermosphere). In section~\ref{sec:regime_three_requirements}, we describe the requirements for regime three. In section~\ref{sec:regime_three_atmospheric_model}, we show how to model the conducting profile of the thermosphere to find the conditions at the Bondi radius, with which the XUV-induced mass loss rate is found. In section~\ref{sec:diffusion_limit}, we describe a procedure for modeling mass loss during regime three, in which the loss of hydrogen is limited by diffusion through the atmosphere.

\subsection{Regime three requirements}
\label{sec:regime_three_requirements}

The temperature profile of a chemically homogeneous atmosphere composed of an ideal gas is given by the most efficient heat flow mechanism. Above the photosphere, there is a steep temperature gradient because energy cannot be efficiently lost through radiation (the gas is optically thin) or convection (the gas is stably stratified if it is gravitationaly bound), so conduction is the only major heat loss mechanism available \citep{Spitzer1949,Chamberlain1962,Gross1972}. A planet will therefore transition from regime one or two to regime three when its Bondi radius is equal to or larger than the photosphere. This will occur when the optical depth at the Bondi radius is equal to 2/3 (see equation~\ref{eq:tau_B}). As explained previously, this condition can be expressed as a Bondi radius temperature,
\begin{equation}
    T_{\rm B} \leq \frac{6 \kappa_{\rm th} G M_{\rm n}\bar{\mu}\rho_{\rm B}}{\gamma^{2}k_{\rm B}},
\end{equation}
with the corresponding basal atmospheric temperature, depending on the mass of the nucleus and atmosphere, and the thermodynamic conditions of the interior.

\subsection{Atmospheric model}
\label{sec:regime_three_atmospheric_model}

\begin{figure}[htbp]
    \centering
    \includegraphics[scale = 0.3]{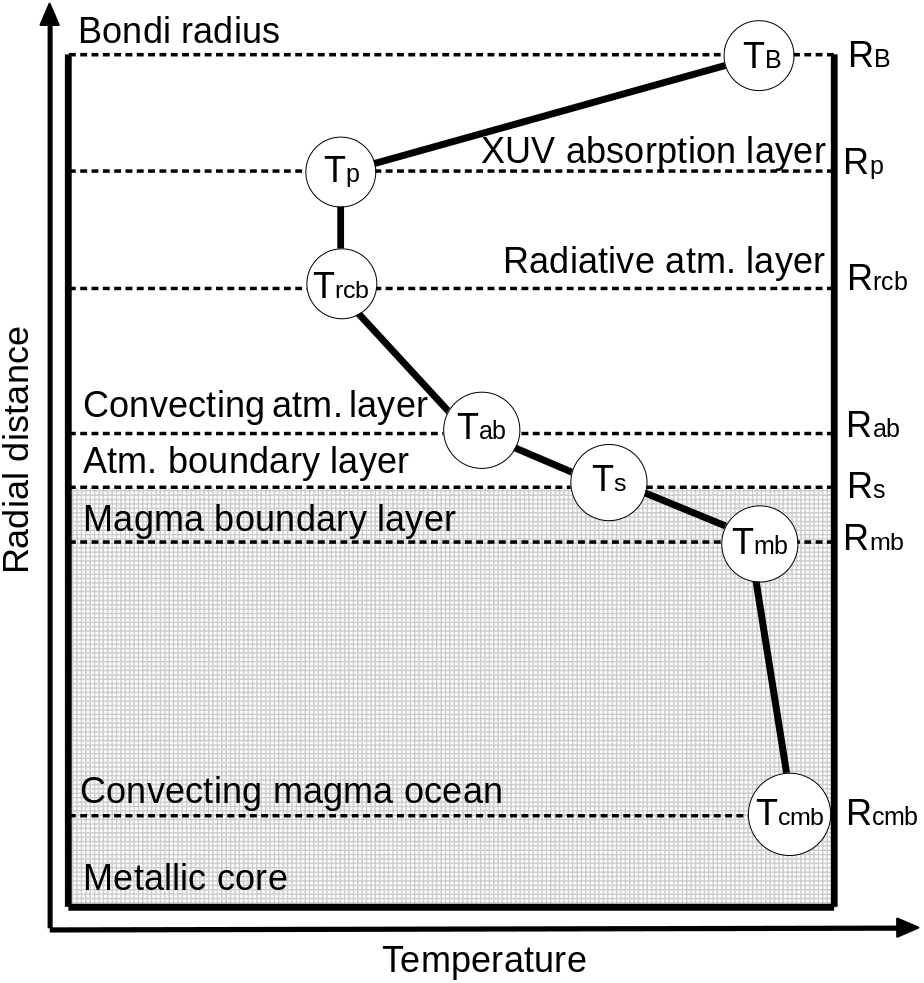}
    \caption{The expected temperature profile of the planet during regime three. Temperature and distance are not to scale; $T_{\rm B}$, $T_{\rm p}$, $T_{\rm rcb}$, $T_{\rm ab}$, $T_{\rm s}$ $T_{\rm mb}$, and $T_{\rm cmb}$ are the temperatures at the Bondi radius, photosphere, radiative-convective boundary, top of the atmospheric boundary layer, surface of the nucleus, top of the convecting magma ocean, and metallic core-magma ocean boundary layer, respectively. $R$ with the relevant subscript marks the radius at which each temperature is defined.}
    \label{fig:regime_three}
\end{figure}

\begin{figure}[ht]
    \centering
    \includegraphics[scale = 0.3]{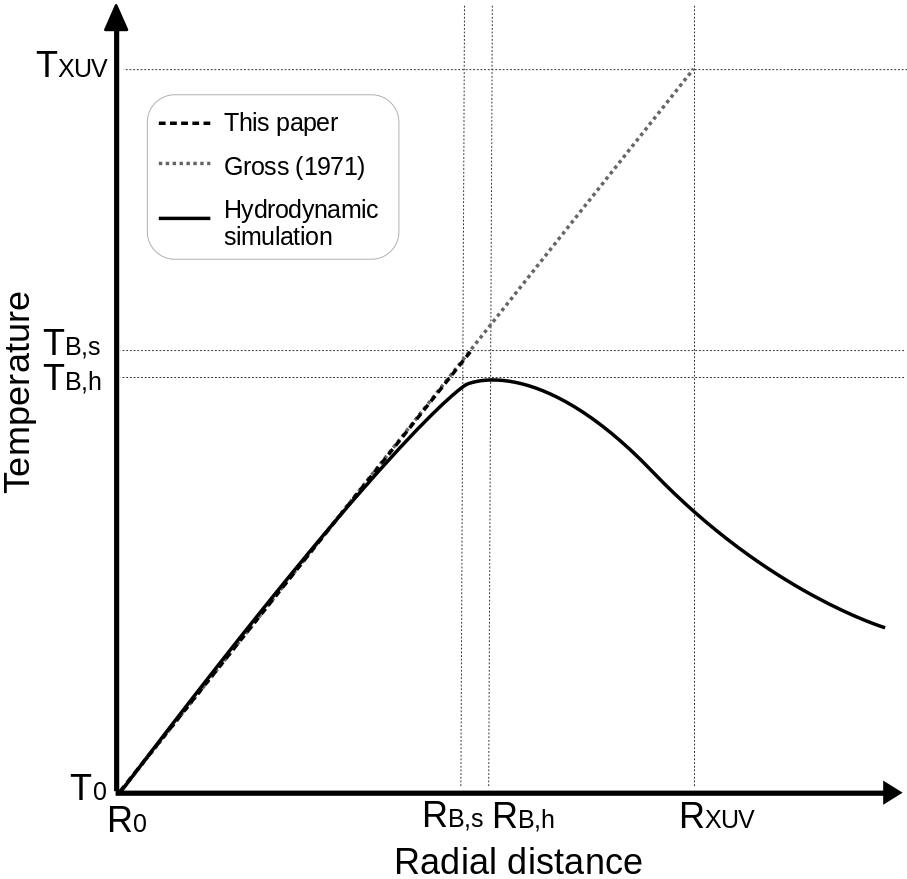}
    \caption{Schematic drawing showing the difference between the model of \citet{Gross1972} (gray dotted line), the typical results from hydrodynamic models (black solid line), and our approach (black dashed line). $T_{0}$ and $R_{0}$ are the bottom reference temperature and radius, $R_{\rm B,s}$ and $T_{\rm B,s}$ are the Bondi radius and temperature predicted by our approach, $R_{\rm B,h}$ and $T_{\rm B,h}$ are the Bondi radius and temperature predicted by hydrodynamic models, and $R_{\rm XUV}$ and $T_{\rm XUV}$ are the radius and temperature at the XUV absorption region respectively. The conditions at the Bondi radius ($R_{\rm B}$) define the mass loss rate, so we do not consider any altitudes above it in our theory.}
    \label{fig:schematic_temperature}
\end{figure}

In this regime, atmospheric evaporation is driven by X-ray and ultraviolet irradiation. The region above the photosphere is called the thermosphere and can be modeled in two ways: as a static gas \citep{Chamberlain1962,Gross1972,Horedt1982} or a hydrodynamic one \citep{Sekiya1980,Watson1981,Zahnle1986,Yelle2004,Lammer2013}. The static approach has been shown to adequately model the thermospheric profile of Earth \citep{Bates1951,Bates1959,Kelly1983,Hedin1991,Gusev2006,Emmert2021}, but it predicts temperatures that are implausibly high when applied to primordial atmospheres \citep{Gross1972,Horedt1982}. As first proposed by \citet{Opik1963}, if the thermospheric temperatures exceed the critical temperature $T {=} 2GM_{\rm n}\bar{\mu}{/}\left(\gamma k_{\rm B}r\right)$, the atmosphere becomes transonic so that the static model is no longer applicable all the way to the XUV absorption radius $R_{\rm XUV}$. It has therefore become common to model the thermospheres of exoplanets with the hydrodynamic approach, predicting significantly lower temperatures \citep[e.g.,][]{Kubyshkina2018(1),Kubyshkina2018(2)}. In this paper we adopt the static conduction model of \citet{Gross1972}, but only until the Bondi radius where gases become hydrodynamic. Beyond the Bondi radius, gases cool from their adiabatic free expansion, so our atmospheric model has its highest temperatures at the Bondi radius. This is a different approach from the one adopted by \citet{Gross1972}, who assumes that the temperature keeps increasing until the XUV-absorption radius $R_{\rm XUV}$. Figure~\ref{fig:schematic_temperature} shows a schematic drawing of our approach when compared with \citet{Gross1972} and a standard hydrodynamic approach.

We adopt a conductive temperature profile for the thermosphere,
\begin{equation}
    \frac{Q_{\rm XUV}-L}{4 \pi r^{2}} = k\frac{dT}{dr},
\label{eq:conduction_model_1}
\end{equation}
where the left-hand side describes the energy balance with $Q_{\rm XUV}$ being the incoming XUV luminosity, $L$ the cooling rate, and the denominator of the fraction being the area; the right-hand side is the conduction equation. The above equation assumes that the heat flux through each layer of the upper atmosphere is constant. This assumption is justified because the energy contribution from XUV irradiation vanishes with optical depth as follows \citep{Sekiya1980},
\begin{equation}
    \frac{F_{\rm XUV}\left(\tau\right)}{F_{\rm XUV}\left(0\right)} \simeq \frac{\exp{\left(-\tau\right)}}{1+2\tau}, 
\end{equation}
where $\tau$ scales almost linearly with distance. In other words, the energy contribution from absorbed XUV photons below the Bondi radius scales as ${\in} \mathcal{O}\left[\exp{\left(-r\right)}\right]$ whereas the energy contribution from conduction from the Bondi radius downwards scales as ${\in} \mathcal{O}\left(1/r\right)$. In addition, \citet{Gross1972} showed that cooling is negligible because gas below the XUV absorption radius $R_{\rm XUV}$ is optically thick to XUV photons, so the reemission of an XUV photon will probably result in a subsequent reabsorption elsewhere in the vicinity. Setting $L {=} 0$ and expressing $Q_{\rm XUV}$ as $\pi R_{\rm B}^{2} \xi_{\rm XUV}F_{\rm XUV}$ transforms equation~\ref{eq:conduction_model_1} to
\begin{equation}
    \xi_{\rm XUV} \pi R_{\rm B}^{2}F_{\rm XUV} = 4 \pi r^{2} k\frac{dT}{dr},
\label{eq:conductive_profile}
\end{equation}
where $\xi_{\rm XUV}$ is an efficiency factor accounting for the fraction of incoming XUV photons that get absorbed and heat the local gas. The thermal conductivity can be expressed as a function of the temperature $C_{1} T^{C_{2}}$, where $C_{1}$ and $C_{2}$ are constants (see equation~\ref{eq:chapman_conductivity}), so
\begin{equation}
    \frac{dT}{dr}=\frac{\xi_{\rm XUV} F_{\rm XUV} R_{\rm B}^{2}}{4 C_{1} T^{C_{2}}r^{2}}.
\label{eq:conductive_temp_diff}
\end{equation}
The efficiency factor $\xi_{\rm XUV}$ may be expressed as a function of two components:
\begin{equation}
    \xi_{\rm XUV} = \xi_{2} \xi_{3},
\end{equation}
where $\xi_{2}$ is the absorption efficiency of XUV energy by the gas. This is found by solving the following integrals,
\begin{equation}
    \xi_{2} = \frac{\int_{\rm XUV} \xi(E) \mathcal{S}(E,t)dE}{\int_{\rm XUV} \mathcal{S}(E,t)dE},
\end{equation}
where $\xi(E)$ is the heating efficiency of an ionizing photon with energy $E$ \citep{Dalgarno1999}, and $\mathcal{S}(E,r)$ is the shape of the stellar XUV spectrum as a function of energy and time \citep{Locci2019}. The time dependency of the spectrum is estimated with
\begin{equation}
    \mathcal{S}(E,t) = \mathcal{C}_{\rm so}(t)\mathcal{S}_{\rm so}(E) + \mathcal{C}_{\rm ha}(t)\mathcal{S}_{\rm ha}(E),
\end{equation}
where $\mathcal{S}_{\rm so}(E)$ and $\mathcal{S}_{\rm ha}(E)$ are the soft and hard components of the XUV bands, respectively, and $\mathcal{C}_{\rm so}$ and $\mathcal{C}_{\rm ha}$ are coefficients that evolve with time \citep{Micela2002}. Using the spectral data from \citet{Raymond1977}, \citet{Locci2018,Locci2019} solved the above integral and found that $\xi_{2} {\simeq} 0.8$.

\begin{figure}[htbp]
    \centering
    \includegraphics[scale = 0.9]{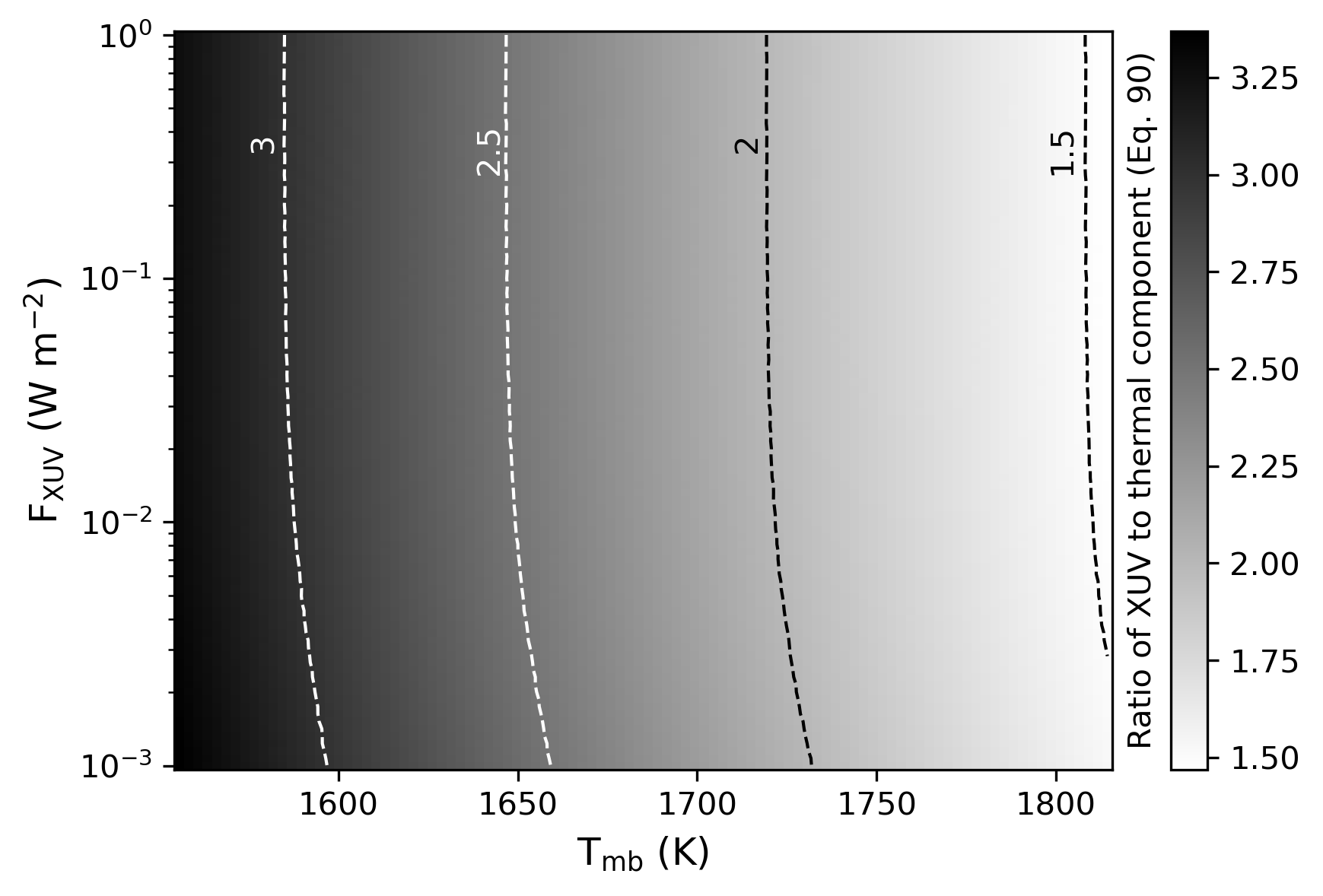}
    \caption{The ratio of the XUV component divided by the thermal component of equation~\ref{eq:Bondi_limited_XUV}. The ratio approaches one at higher magma ocean temperatures, and then becomes zero when the system transitions into regime two.}
    \label{fig:ratio_components}
\end{figure}

The parameter $\xi_{3}$ accounts for the fraction of the deposited energy that is used in the dissociation of atomic and molecular hydrogen. We define $\xi_{3}$ as
\begin{equation}
    \xi_{3} = \frac{h\bar{f} - \Delta H}{h\bar{f}},
\end{equation}
where $h$ is Planck's constant, $\bar{f}$ is the average frequency of an XUV photon ($h\bar{f}{\approx} 20~{\rm eV}$), and $\Delta H$ is the enthalpy of ionization. The extreme ultraviolet (EUV) absorption cross-section of hydrogen is three orders of magnitude higher than that for X-ray photons \citep[e.g.,][]{Spitzer1978}, so EUV photons contribute more to heating at the highest sections of the thermosphere. X-ray photons would, however, penetrate deeper into the atmosphere and heat the regions closer to the photosphere more than EUV photons \citep{Kubyshkina2018(1)}. We do not include the energy dilution from the absorption of X-rays by heavier species because our model assumes primordial atmospheres with no further atmospheric enrichment (see sections~\ref{sec:intro} and \ref{sec:cloud_formation}). We therefore include ionization arising from both energy bands. We consider the ionization of atomic hydrogen only; incorporating molecular hydrogen, photochemistry, other chemical impurities, and the possibility of non-thermodynamic equilibrium \citep{Dewan1961,Fridman2008} are beyond the scope of this paper. The total enthalpy of an atomic hydrogen plasma is \citep{Capitelli2008}
\begin{equation}
    H = \frac{5}{2}\left(1+X\right)k_{\rm B}T+\left(1-X\right)E_{\rm H}+XI_{\rm H},
\end{equation}
where $X$ is the ionization degree, $I_{\rm H}$ is the ionization energy of atomic hydrogen, $E_{\rm H}$ is the electronic energy of atomic hydrogen
\begin{equation}
    E_{\rm H} = I_{\rm H}\left(1-\frac{1}{\mathcal{N}^{2}}\right),
\end{equation}
and $\mathcal{N}$ is the principal quantum number. The enthalpy of nonionized gas at zero Kelvin and at a finite temperature are
\begin{equation}
    H_{\rm ni}\left(0\right) = E_{\rm H}
\end{equation}
and
\begin{equation}
    H_{\rm ni}\left(T\right) = \frac{5}{2}k_{\rm B}T+E_{\rm H},
\end{equation}
respectively, so
\begin{equation}
    H_{\rm ni}\left(T\right) = H_{\rm ni}\left(0\right) + \frac{5}{2}k_{\rm B}T.
\end{equation}
By performing a similar manipulation for fully ionized gas (subscript i), one arrives at
\begin{equation}
    H_{\rm i}\left(T\right) = H_{\rm i}\left(0\right) + 5k_{\rm B}T - \Delta I_{\rm H},
\end{equation}
where $\Delta I_{\rm H}$ is an ad hoc correction that accounts for the Coulomb interaction of other ions and electrons \citep{Griem1962,Capitelli1970,Capitelli2008,Zaghloul2008,Zaghloul2009}. The enthalpy change from ionization is 
\begin{equation}
    \Delta H\left(T\right) = \Delta H\left(0\right) + \frac{5}{2}k_{\rm B}T - \Delta I_{\rm H},
\end{equation}
where
\begin{equation}
    \Delta H\left(0\right) = I_{\rm H}-E_{\rm H}.
\end{equation}
Assuming that before ionization all gas particles are in their ground state, that is, the principle quantum number is one, we have $E_{\rm H}{=}0$ and $\Delta H\left(0\right){=} I_{\rm H} {=} 2.178{\times}10^{-18}~{\rm J}$ (13.595 eV). The variable $\Delta I_{\rm H}$ is given by \citep{Griem1962},
\begin{equation}
    \Delta I_{\rm H} = 2\left(Z_{j}+1\right)q_{\beta}^{3}\left(\frac{\pi}{k_{\rm B}T}\right)^{\frac{1}{2}}\left(n_{\beta}+\sum^{j = N}_{j=1} Z_{j}^{2}n_{j}\right)^{\frac{1}{2}} \left(\frac{1}{\varepsilon_{0}\varepsilon_{\rm r}}\right)^{\frac{3}{2}},
\label{eq:dI}
\end{equation}
where $Z_{j}$ is the atomic number, $q_{\beta}$ is the charge of an electron, $n$ is the number density, $N$ is the number of species present (i.e., one for a pure gas), $\varepsilon_{0}$ is the permittivity of free space, and $\varepsilon_{\rm r}$ is the relative permittivity. For atomic hydrogen $N {=} 1$, $Z_{1} {=} 1$ and $\varepsilon_{\rm r} {\simeq} 1$. Through further manipulation, equation~\ref{eq:dI} becomes
\begin{equation}
    \Delta I_{\rm H} \simeq 52I_{\rm H}\left(\frac{XP}{T^{2}}\right)^{\frac{1}{2}},
\label{eq:I_reduction}
\end{equation}
which can be shown to always be negligible by considering the lower and upper bounds of the thermosphere at the photosphere and Bondi radius, respectively. At the photosphere, the XUV optical depth is very high, so the degree of ionization $X$ is dictated by thermal collisions, which is adequately modeled by the Saha equation \citep{Saha1920,Saha1921,Fridman2008}. For typical photospheric conditions, the degree of ionization is vanishingly small, so $\Delta I_{\rm H}$ is negligible. At the Bondi radius, ionization occurs because of incoming XUV photons so that $X$ approaches one, though pressures are very low and temperatures are high, so $\Delta I_{\rm H}$ is also negligible. $\Delta I_{\rm H}$ is therefore several orders of magnitude smaller than the ionization energy of hydrogen and is thus ignored. The efficiency is therefore
\begin{equation}
\begin{split}
    \xi_{3} &= \frac{h\bar{f} - I_{\rm H}-\frac{5}{2}k_{\rm B}T}{h\bar{f}} \\
    & = \frac{T_{1}-T}{T_{2}},
\end{split}
\label{eq:efficiency_3}
\end{equation}
so
\begin{equation}
    \xi_{\rm XUV} = \xi_{2} \frac{T_{1}-T}{T_{2}},
\label{eq:efficiency_XUV}
\end{equation}
where $T_{1}$ and $T_{2}$ are constants of value $\rm 2.97{\times}10^{4}~{\rm K}$ and $\rm 9.28{\times}10^{4}~K$, respectively. Equation~\ref{eq:efficiency_XUV} sets an upper limit for the mass loss efficiency because it does not consider the absorption of XUV photons by escaping winds. This effect is, however, probably minor because escaping winds are mostly composed of ionized atomic hydrogen \citep[i.e., free protons;][]{Kubyshkina2018(2)} that cannot absorb photons. Free protons cannot absorb photons because it would violate energy and momentum conservation. Equation~\ref{eq:conductive_temp_diff} can therefore be expressed as
\begin{equation}
    \frac{dT}{dr}= \frac{T_{1}-T}{T_{2}}\frac{\xi_{2} F_{\rm XUV} R_{\rm B}^{2}}{4 C_{1} T^{C_{2}}r^{2}}.
\label{eq:conductive_temp_diff_2}
\end{equation}
Integrating the above equation with the lower and upper limits at the photosphere and Bondi radius, respectively, leads to the following:
\begin{equation}
\begin{split}
    \frac{T_{2}}{\left(C_{2}+1\right)T_{1}}\left[T^{C_{2}+1}_{B} {}_{2}{\rm F}_{1}\left(1,C_{2}+1;C_{2}+2;\frac{T_{\rm B}}{T_{1}}\right)\right. &- T^{C_{2}+1}_{\rm p} \left.{}_{2}{\rm F}_{1}\left(1,C_{2}+1;C_{2}+2;\frac{T_{\rm p}}{T_{1}}\right)\right] \\ &=\frac{\xi_{2}F_{\rm XUV}}{4 C_{1}}\left(\frac{2GM_{\rm n}\bar{\mu}_{\rm i}}{\gamma k_{\rm B}T_{\rm B}}\right)^{2}\left(\frac{1}{R_{\rm p}} - \frac{\gamma k_{\rm B}T_{\rm B}}{2GM_{\rm n}\bar{\mu}_{\rm i}}\right),
\end{split}
\label{eq:temperature_solution}
\end{equation}
where $\bar{\mu}_{\rm i}$ is the ionized mean molecular weight ($\rm 0.5~amu$) and ${}_{2}{\rm F}_{1}$ is the hypergeometric function, which does not deviate much from unity. It is therefore possible to approximate ${}_{2}{\rm F}_{1}$ with the first two terms of its power series so that
\begin{equation}
\begin{split}
    \frac{T_{2}}{\left(C_{2}+1\right)T_{1}}\left[T^{C_{2}+1}_{\rm B} \left(1+\frac{C_{2}+1}{C_{2}+2}\frac{T_{\rm B}}{T_{1}}\right)\right. &- T^{C_{2}+1}_{\rm p} \left.\left(1+\frac{C_{2}+1}{C_{2}+2}\frac{T_{\rm p}}{T_{1}}\right)\right] \\ &=\frac{\xi_{2}F_{\rm XUV}}{4 C_{1}}\left(\frac{2GM_{\rm n}\bar{\mu}_{\rm i}}{\gamma k_{\rm B}T_{\rm B}}\right)^{2}\left(\frac{1}{R_{\rm p}} - \frac{\gamma k_{\rm B}T_{\rm B}}{2GM_{\rm n}\bar{\mu}_{\rm i}}\right).
\end{split}
\label{eq:temperature_solution_2}
\end{equation}
Equation~\ref{eq:temperature_solution_2} can be solved through iteration. The density profile is found by combining equation~\ref{eq:conductive_temp_diff_2} with the equation for hydrostatic equilibrium:
\begin{equation}
    \frac{dT}{dP}= -\frac{T_{1}-T}{T_{2}}\frac{\xi_{2} F_{\rm XUV} R_{\rm B}^{2} k_{\rm B} T^{1-C_{2}}}{4 C_{1} G M_{\rm n} \mu P},
\end{equation}
which can be solved to give
\begin{equation}
\begin{split}
    \rho_{\rm B} = &\frac{2GM_{\rm n}\bar{\mu}}{3\kappa_{\rm th} k_{\rm B}T_{\rm B} R_{\rm p}^{2}} \exp{\left\{-\frac{4 C_{1} GM_{\rm n}\bar{\mu}_{\rm i} T_{2}}{\varepsilon_{2}C_{2}F_{\rm XUV} k_{\rm B} T_{1}} \left[\frac{\gamma k_{\rm B} T_{\rm B}}{2GM_{\rm n}\bar{\mu}_{\rm i}}\right]^{2}\left[T_{\rm B}^{C_{2}} \left(1+\frac{C_{2}}{C_{2}+1} \frac{T_{\rm B}}{T_{1}} \right)\right.\right.} \\ &{\left.\left.-T_{\rm p}^{C_{2}} \left(1+\frac{C_{2}}{C_{2}+1} \frac{T_{\rm p}}{T_{1}} \right) \right] \right\}}.
\end{split}
\label{eq:density_solution}
\end{equation}
The hypergeometric function was again replaced by the first two terms of its power approximation, and the pre-exponential parameters and numbers come from the definition of the photosphere (see appendix). Because the conditions at the Bondi radius are now known, the mass loss rate can be estimated with
\begin{equation}
    \dot{M}_{\rm B} \approx \left(\pi \xi_{1} R^{2}_{\rm B} \rho_{\rm B} u_{\rm B}\right)_{\rm XUV} + \left(3\pi \xi_{1} R^{2}_{\rm B} \rho_{\rm B} u_{\rm B}\right)_{\rm th}.
\label{eq:Bondi_limited_XUV}
\end{equation}
Equation~\ref{eq:Bondi_limited_XUV} assumes that one fourth of the planet's surface area is exposed to XUV irradiation whereas the other three fourths experience mass loss due to thermal energy. Figure~\ref{fig:ratio_components} shows the ratio of the XUV component divided by the thermal component of equation~\ref{eq:Bondi_limited_XUV}.

Our strategy for modeling the thermal evolution of the planet during regime three (with and without the diffusion limit) is identical to that of regime two. After setting the initial magma ocean temperatures, we use equations~\ref{eq:flux_through_magma} and \ref{eq:temperature_through_atmosphere} to find the temperature at the base of the atmosphere (i.e., above the top boundary layer of the magma and bottom boundary layer of the atmosphere). Using equation~\ref{eq:F_rcb_complete}, we find the outward heat flux, which is then used in equation~\ref{eq:interior_cooling} to track planetary cooling.

\subsection{Diffusion limited loss}
\label{sec:diffusion_limit}

XUV-induced photoevaporation preferentially removes hydrogen relative to other species because of its low mass. Heavier species are lost by gaining escape velocity through continuous collisions with the lighter hydrogen \citep{Hunten1987,Zahnle1990,Chassefiere1996,Luger2015}. The preferential loss of hydrogen occurring at the Bondi radius will briefly generate a local compositional gradient that will trigger the diffusive transport of hydrogen from the deeper layers. Diffusion would be the only mechanism that can restock the locally depleted hydrogen because the thermosphere is stably stratified. The above prescription would indicate that there must exist an equilibrium between the diffusive transport of hydrogen and its loss at the Bondi radius. In the early stages of regime three when the internal temperatures can still be high, eddy diffusion will be the dominant form of mass transport above the photosphere, whereas molecular diffusion would dominate if a heterosphere forms after sufficient cooling.

\subsubsection{Eddy diffusion limit}
\label{sec:eddy_diffusion}

The eddy diffusion limit applies when the molecular diffusion coefficient, $D$, is smaller than the eddy diffusion coefficient, $K_{\rm zz}$, at every location above the photosphere. Whereas the molecular diffusion coefficient is well constrained, the eddy diffusion coefficient is uncertain. Above Earth's tropopause, mixing occurs because of gravitational waves breaking and releasing potential energy that leads to mechanical mixing \citep[e.g.,][]{Lindzen1971,Lindzen1981}. \citet{Lindzen1971} suggested that wave breaking should result in an eddy diffusion coefficient that scales with the square root of the inverse pressure.
\citet{Parmentier2013} found the same pressure dependency from their 3-D simulations of exoplanet atmospheres, whereas \citet{Charnay2015} proposed that an exponent of $-2/5$ produced more accurate results. They suggest that in close orbiting exoplanets, the strong temperature contrast between the dayside and nightside gives rise to global horizontal winds that also contribute to mixing. For a convecting system, the eddy diffusion coefficient scales with $-1/3$ (see equation~\ref{eq:Kzz_convection} below) whereas it scales with $-1/2$ for gravitational wave breaking, explaining the intermediate value of $-2/5$ used in their model. On Earth, the eddy diffusion coefficient increases until the location where gravitational waves break, after which the pressure trend of the eddy diffusion coefficient reverses \citep[e.g.,][]{Shimazaki1971,Kirchhoff1983,Lubken1997,Vlasov2015}. Because horizontal convection also contributes to mixing on hot exoplanets, it is uncertain if a reversal will occur, so we adopt the eddy diffusion relation proposed by \citet{Charnay2015}
\begin{equation}
    K_{\rm zz} = K_{\rm rcb}\left(\frac{P}{P_{\rm rcb}}\right)^{-\frac{2}{5}},
\label{eq:Kzz_gravity}
\end{equation}
where $K_{\rm rcb}$ is the eddy diffusion coefficient at the radiative-convective boundary given by \citep{Gierasch1985,Ackerman2001,Lupu2014}
\begin{equation}
    K_{\rm rcb} \approx 0.01H_{\rm rcb}\left(\frac{\lambda}{H_{\rm rcb}}\right)^{\frac{4}{3}}\left(\frac{F_{\rm rcb}}{\rho_{\rm rcb}}\right)^{\frac{1}{3}},
\label{eq:Kzz_convection}
\end{equation}
and $\lambda$ is the mixing length. Equation~\ref{eq:Kzz_convection} is derived by assuming that gases are inviscid and well described by mixing length theory. Because the mixing length is poorly constrained, it is generally assumed to be equal to the scale height. There are, however, two major issues with this assumption. First, the scale height and mixing length are two separate concepts, and second, they have different dependencies. A more reasonable approach may be to use the scaling arguments provided by atmospheric models, or to apply Kolmogorov theory. According to the atmospheric models of \citet{Smith1998} and \citet{Charnay2015}, the mixing length is approximately one tenth of the scale height. A similar result is attained from Kolmogorov theory \citep{Kolmogorov1941(1),Kolmogorov1941(2)}, stating that most energy transfer (and hence most mass transfer) is from eddies that are of the order one tenth the length scale \citep[i.e., the energy-containing range;][]{Pope2000}. The length scale is comparable to the scale height because coherency can only be maintained in relatively isobaric regions. The molecular diffusion coefficient is given by Chapman-Enskog theory \citep{Chapman1970}, and it scales as
\begin{equation}
    D = D_{\rm rcb}\left(\frac{T}{T_{\rm rcb}}\right)^{\frac{3}{2}}\left(\frac{P}{P_{\rm rcb}}\right)^{-1},
\end{equation}
where pressure can change by many orders of magnitude whereas the temperature changes by at most one, so the molecular diffusion coefficient can also be approximated as only depending on the pressure
\begin{equation}
    D \approx D_{\rm rcb}\left(\frac{P}{P_{\rm rcb}}\right)^{-1}.
\end{equation}
Here, $D_{\rm rcb}$ is the molecular diffusion coefficient at the radiative-convective boundary given by \citep{Chapman1970}
\begin{equation}
    D_{\rm rcb} \approx \frac{3k_{\rm B}T_{\rm rcb}}{8P_{\rm rcb}d^{2}}\sqrt{\frac{k_{\rm B}T_{\rm rcb}}{2\pi \mu_{\rm H}}}.
\end{equation}
For the eddy diffusion limit to apply, it is necessary for the eddy diffusion coefficient to be greater than the molecular diffusion coefficient at the Bondi radius
\begin{equation}
    K_{\rm B} > D_{\rm B}.
\label{eq:turbopause}
\end{equation}
If the above inequality is satisfied, the eddy diffusion limit can be found by considering the boundary conditions at the photosphere. Below the photosphere, temperatures decrease with height \citep[e.g.,][]{Parmentier2014} whereas above the photosphere they increase due to the conductive temperature profile \citep[e.g,][]{Bates1951,Bates1959}. This suggests that there must exist a boundary where $dT/dr {=} 0$, which is where the diffusion limit is evaluated. We adopt the same formulation as Fick's first law of diffusion but for eddy diffusion at the photospheric boundary
\begin{equation}
    F_{\rm H} = -K_{\rm p}\frac{d\rho}{dr},
\end{equation}
where the temperature is treated as a constant because $dT/dr {=} 0$ at the boundary. By inserting hydrostatic equilibrium one gets
\begin{equation}
\begin{split}
    F_{\rm H} &= K_{\rm p}\frac{\rho_{\rm p} g_{\rm p}\mu_{\rm p}}{k_{\rm B}T_{\rm p}}, \\
    &= K_{\rm p}\frac{\rho_{\rm p}}{\mathcal{H}_{\rm p}},
\end{split}
\end{equation}
with $\mathcal{H}_{\rm p}$ being the scale height at the photosphere. The eddy diffusion limited mass flow is
\begin{equation}
    \dot{M}_{\rm a} = 4 \pi R^{2}_{\rm p} K_{\rm p}\frac{\rho_{\rm p}}{\mathcal{H}_{\rm p}}.
\label{eq:eddy_diffusion_limit}
\end{equation}

\subsubsection{Molecular diffusion limit}

If equation~\ref{eq:turbopause} is not satisfied, molecular diffusion will become the limiting factor for mass loss \citep{Hunten1973,Kasting1983,Zahnle1986,Catling2001,Zahnle2019,Zahnle2020}. The equation for the molecular diffusion limited mass loss is similar to equation~\ref{eq:eddy_diffusion_limit}, but with the eddy diffusion coefficient replaced by the molecular one, as well as all parameter values being given by the conditions at the turbopause (i.e., the location where $K_{\rm zz} {=} D$) and not the photosphere:
\begin{equation}
    \dot{M}_{\rm a} = 4 \pi R^{2}_{\rm t} D_{\rm t}\frac{\rho_{\rm t}}{\mathcal{H}_{\rm t}}.
\label{eq:molecular_diffusion_limit}
\end{equation}

\section{Representative results}

\begin{figure}[htbp]
    \centering
    \includegraphics[width=0.75\columnwidth]{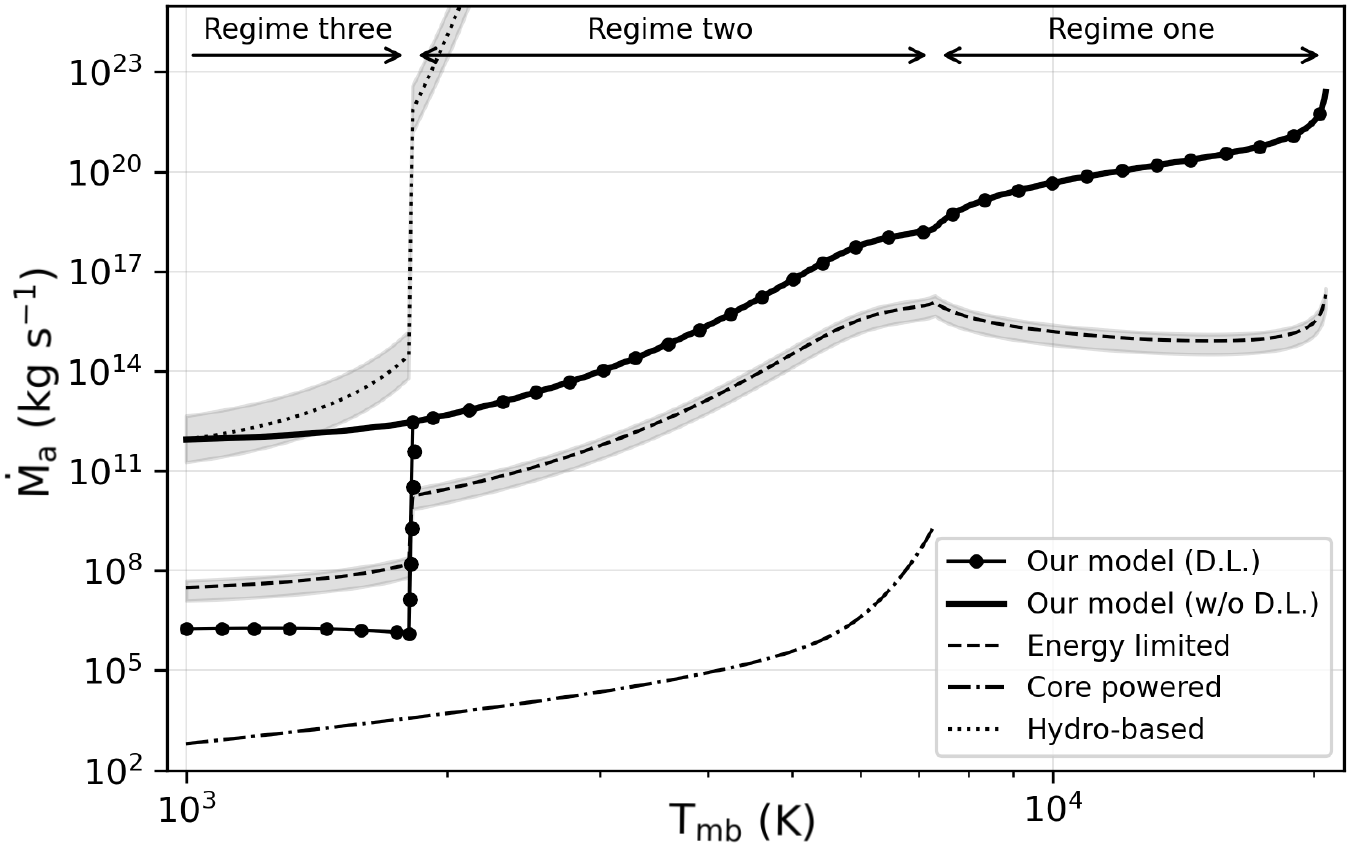}
    \caption{The predicted atmospheric evaporation rates as a function of the surface temperature for an exoplanet that is three times the mass of Earth with a core-mass fraction of 26\%, a primordial (hydrogen-rich) atmosphere that is 1\% of the total planetary mass, and equilibrium temperature of 500 K, experiencing an XUV radiant flux of 0.1$\rm ~W~ m^{-2}$. The black solid with circles, thick black solid, dashed, dash-dotted, and dotted lines are for our model with and without the diffusion limited mass loss included, the energy limited model \citep{Watson1981}, the core powered model \citep{Biersteker2019,Biersteker2021}, and the hydro-based model \citep{Kubyshkina2018(1),Kubyshkina2018(2)}. The gray regions mark the typical uncertainty of the hydro-based and energy limited models. Regime one was modeled with equation~\ref{eq:Bondi_limited}, where the radius, density, and temperature (for calculating the velocity in equation~\ref{eq:velocity}) were given by equations~\ref{eq:Bondi_radius}, \ref{eq:Bondi_density}, and \ref{eq:T_B}, respectively. Regime two uses equations~\ref{eq:rho_regime_two} and \ref{eq:rho_rcb_regime_2} for the density, and \ref{eq:transition_temperature} for the temperature. The mass loss rate in regime three without the diffusion limit (w/o D.L.) was estimated with equation~\ref{eq:Bondi_limited_XUV}, where the density and temperature were given by equations~\ref{eq:density_solution} and \ref{eq:temperature_solution_2}, respectively. The mass loss rate in regime three with the diffusion limit (D.L.) was given by equation~\ref{eq:molecular_diffusion_limit}, with the density and eddy diffusion coefficients given by equations~\ref{eq:photosphere_not_hydro} and \ref{eq:Kzz_convection}, respectively.}
    \label{fig:representative_results}
\end{figure}

Having described our atmospheric, interior, and mass loss framework, we now provide some representative results, which we compare with other approaches in the literature. Figure~\ref{fig:representative_results} shows the predicted atmospheric evaporation rates as a function of the surface temperature for an exoplanet that is three times the mass of Earth with a metallic core-mass fraction of 26\%, covered by a primordial (hydrogen-rich) atmosphere that is 1\% of the total planetary mass with an equilibrium temperature of 500 K and being exposed to an XUV radiant flux of $\rm 0.1~ W~m^{-2}$. Our results (with and without the diffusion limit) are compared with the predictions of the energy limited \citep{Watson1981}, core powered mass loss \citep{Biersteker2019,Biersteker2021}, and hydro-based models \citep{Kubyshkina2018(1),Kubyshkina2018(2)}. The gray regions mark the typical uncertainty of the hydro-based and energy limited models.
As shown by Figure~\ref{fig:representative_results}, our model without the diffusion limit (thick solid line) falls within the uncertainty of the hydro-based model for the low temperature range (regime three). The solid line with the circles in regime three accounts for the eddy diffusion limit discussed in section~\ref{sec:eddy_diffusion}. The molecular diffusion limit does not apply anywhere in the atmosphere because it has not formed a heterosphere. The hydro-based model diverges in regime two because it assumes that mass loss depends always on the photospheric radius to the third or fourth power, which becomes very large and migrates outside the Bondi radius at high temperatures. The hydro-based model is an analytic fit to fluid dynamical simulations of the thermosphere (i.e., above $R_{\rm p}$), with the sections below the photosphere being modeled a priori and coupled through Markov chain Monte Carlo algorithms \citep{Fossati2017}. In other words, the hydro-based model assumes a thermosphere, and it is therefore not applicable to regimes one and two in which a thermosphere does not exist. Because the hydro-based model is designed for regime three, it should not be employed for regimes one and two. 

The energy limited model, 
\begin{equation}
    \dot{M}_{\rm a}=\frac{\pi \xi_{\rm el} R_{\rm p}R_{\rm XUV}^{2}F_{\rm XUV}}{GM_{n}K},
\label{eq:energy_limited_approximation}
\end{equation}
depends linearly on the photospheric radius. It is assumed to have a constant heating efficiency of $\xi_{\rm el} = 0.1{-}0.4$ (shown by the gray uncertainty in Figure~\ref{fig:representative_results}), with $K$ being the reduction factor, which is of the order unity. The parameter $R_{\rm XUV}$ is the XUV absorption radius, which is sometimes assumed to be synonymous with the XUV photosphere \citep[i.e., $\tau_{\rm XUV}=2/3$; e.g.,][]{Murray2009}. This assumption is not always justified because it is valid only when the XUV photosphere lies within the Bondi radius. When an atmosphere is experiencing hydrodynamic outflow, the XUV absorption radius is given by the Bondi radius because it is the highest point in the atmosphere (see section~\ref{sec:regime_3}). Not recognizing this distinction leads to implausibly high thermospheric temperatures \citep{Gross1972,Horedt1982}. The energy limited model predicts mass loss rates lower than ours by several orders of magnitude in regimes one and two. It is, indeed, well documented that the energy limited approach underestimates the mass loss rate for hydrodynamic atmospheres \citep[e.g.,][]{Garcia2007,Lammer2013,Kubyshkina2018(1),Kubyshkina2018(2),Krenn2021} because it does not adequately implement stellar thermal irradiation and interior energy. In regime three, the diffusion of hydrogen limits mass loss, so our model with diffusion predicts lower mass loss rates than the energy limited approach \citep[see also][]{Zahnle2019}.

The core powered mass loss model does not rely on the photospheric radius and instead depends on the Bondi radius or the radiative-convective boundary. \citet{Ginzburg2018} use two equations for modeling mass loss: (1) the Bondi limited mass loss approach (our equation~\ref{eq:Bondi_limited}), and (2) an energy limited approach (not to be confused with the energy limited XUV model, given by equation~\ref{eq:energy_limited_approximation}): 
\begin{equation}
    \dot{M}_{\rm a} = \frac{4\pi R_{\rm rcb}^{3} F_{\rm rcb}}{GM_{\rm n}}.
\label{eq:Bondi_energy_limit}
\end{equation}
Equation~\ref{eq:Bondi_energy_limit} assumes that mass loss is limited by the amount of energy required to do work against gravity in restocking the lost mass at the Bondi radius by transporting gas from the radiative-convective boundary upward. The mass loss rate is estimated by adopting the minimum value of equations~\ref{eq:Bondi_limited} and \ref{eq:Bondi_energy_limit}. The above formulation is hard to justify even if matter needs to be restocked in the upper sections of the atmosphere. When an atmosphere loses mass, it will experience an adiabatic expansion because the mass deficit at the Bondi radius will cause hydrostatic equilibrium to no longer be satisfied. The atmosphere will decompress, by using potential energy to do work against gravity in transporting mass to the Bondi radius. The decompressed atmosphere will then assume its new equilibrium thermal structure. This process is independent of the mass of the nucleus because all the energy will come from the decompression of the atmosphere. Energy cannot come from the nucleus because the atmosphere is optically thick (see Figure~\ref{fig:ratio}). The core powered mass loss model predicts lower mass loss rates than our model \citep[i.e.,][]{Ginzburg2018,Biersteker2019,Biersteker2021} because of the adoption of this upper limit.

\begin{figure}[htbp]
    \centering
    \includegraphics{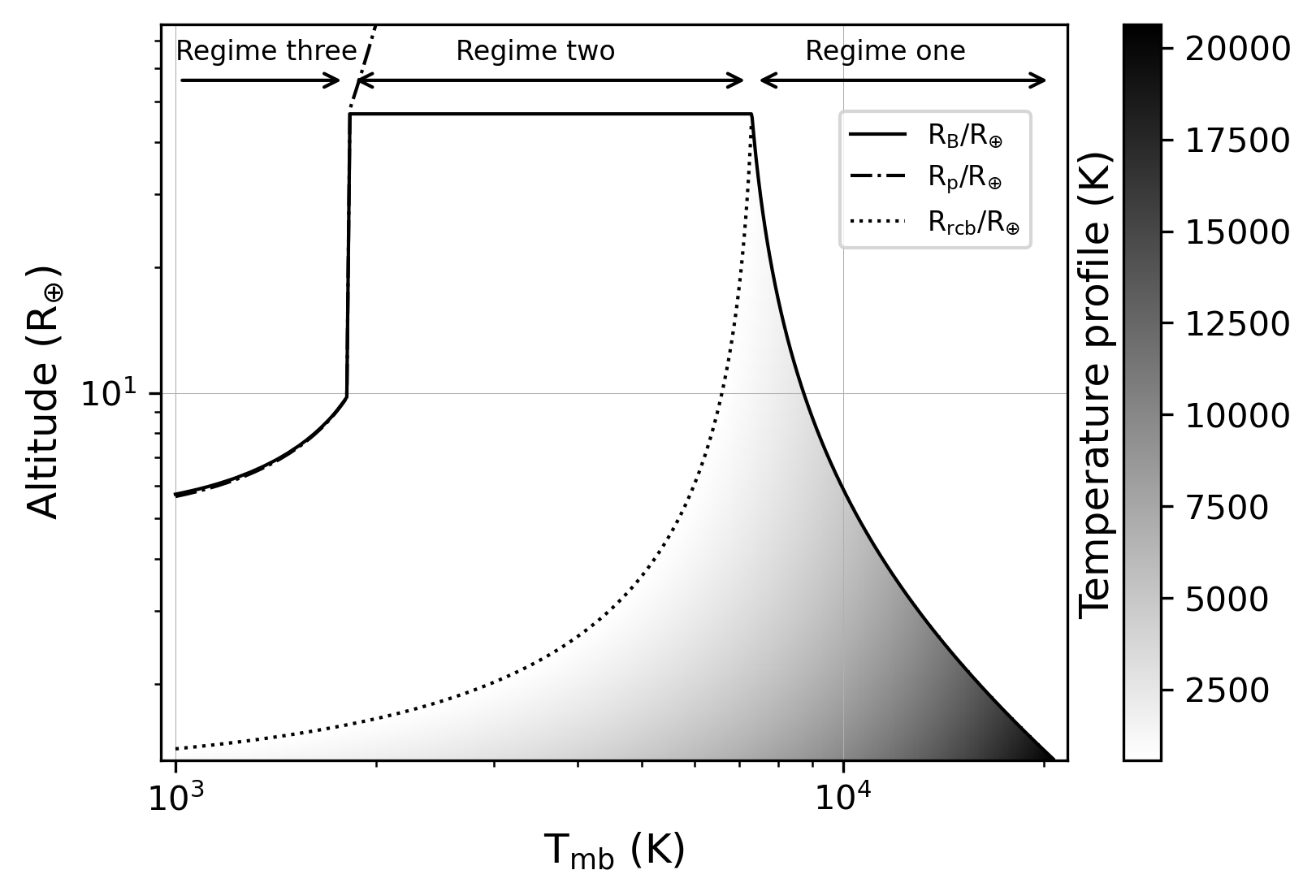}
    \caption{2-D plot showing how the Bondi radius and temperature profile of a planet, with the same properties as those given in Figure~\ref{fig:representative_results}, change with surface temperature. The solid, dash-dotted, and dotted lines are for the Bondi radius, photosphere, and radiative-convective boundary, respectively. The darker area encased between the dash-dotted and solid lines in regime three is the thermosphere.}
    \label{fig:heatmap}
\end{figure}

\begin{figure}[htbp]
    \centering
    \includegraphics[width=0.95\columnwidth]{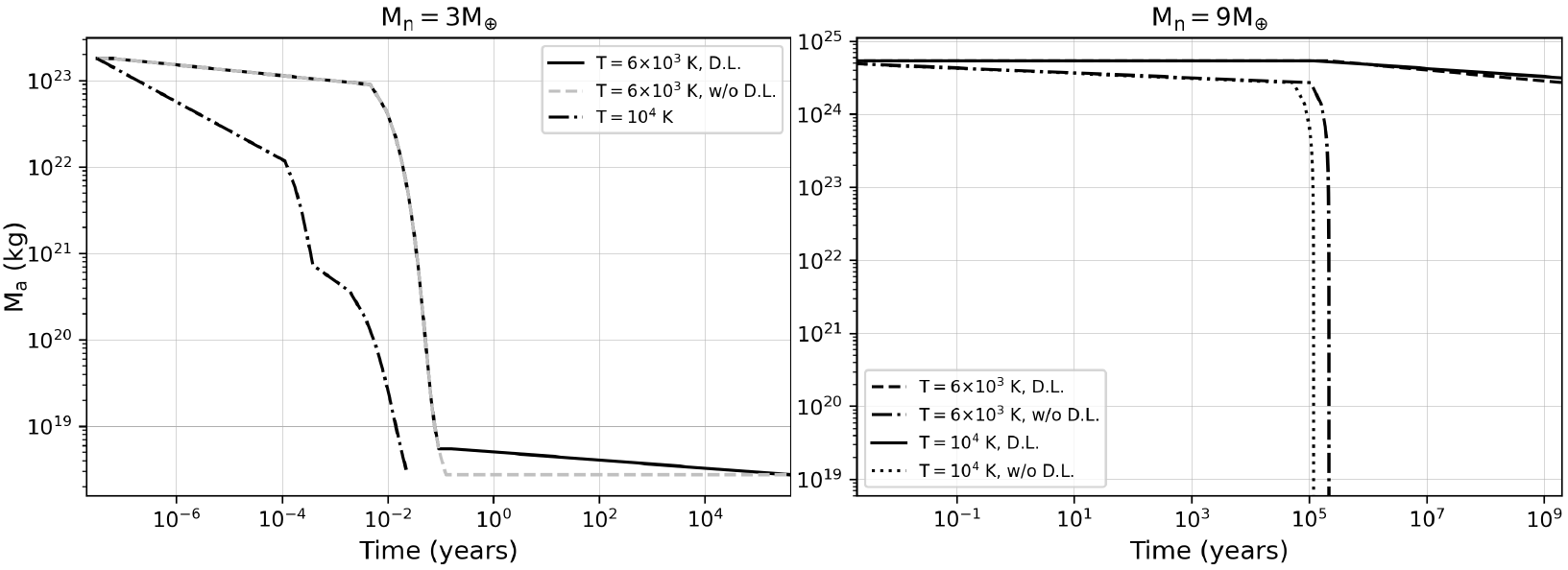}
    \caption{The atmospheric evolution of three and nine Earth mass planets with different internal thermodynamic properties. Left: The black solid, gray dashed, and black dashdotted lines are for $\rm 6000~K$ with diffusion limit, $\rm 6000~K$ without diffusion limit, and $\rm 10,000~K$ with no difference between the diffusion and no diffusion limit cases respectively. Right: The dashed, dashdotted, solid, and dotted lines are for $\rm 6000~K$ with and without the diffusion limit, and $\rm 10,000~K$ with and without the diffusion limit, respectively.}
    \label{fig:time_evolution}
\end{figure}

The applicability of the hydro-based and energy limited models is restricted to regime three, whereas the core powered mass loss model is better suited for regime two. The above mentioned models do not consider the different regimes that an atmosphere assumes under different thermodynamic conditions. Figure~\ref{fig:heatmap} shows how the radius and temperature profile of a planet with a primordial atmosphere (with the planetary properties equal to those used in Figure~\ref{fig:representative_results}) change as a function of the surface temperature, as well as illustrating the three regimes of atmospheric evaporation.
In regime one, the atmosphere is very hot and fully convecting, so the Bondi radius is small. In regime two, a radiative section has formed, but the atmosphere is still optically thick, so XUV irradiation cannot efficiently heat the atmosphere. The atmosphere therefore assumes the equilibrium temperature set by stellar irradiation, resulting in a large Bondi radius. In regime three, the photosphere has migrated beneath the Bondi radius, so the top sections of the atmosphere are optically thin, and XUV heating becomes important. High thermospheric temperatures lead to a contraction of the Bondi radius because gases become hydrodynamic at lower altitudes. The regime a planet is in depends mainly on the mass of the nucleus and atmosphere, as well as the interior temperatures. However, because mass loss and radiative cooling do not occur concurrently (the former is greater than the latter as shown in Figure~\ref{fig:ratio}), a planet is unlikely to transition from regime one to two but rather from regime one to three or from regime two to three. Figure~\ref{fig:time_evolution} shows the atmospheric evolution of three and nine Earth mass planets with different internal thermodynamic properties, respectively. Indeed, the three Earth mass planets experience total atmospheric loss in short time frames, whereas the nine Earth mass planets can survive for billions of years when the diffusion limit is considered. In other words, planets with greater masses are more likely to keep their primordial atmospheres even after their highly energetic formations and exposure to high energy irradiation from their host stars than smaller mass planets. 
The atmospheric mass evolution shown in Figure~\ref{fig:time_evolution} is calculated as follows. Starting with the mantle potential temperature, $T_{\rm mb}$, the temperature at the bottom of the convecting atmosphere $T_{\rm ab}$ is found. The Bondi radius properties, such as the density and temperature, are then evaluated depending on the planet's regime. The mass loss rate and cooling rate are then found, with which the conditions of the nucleus and atmosphere are updated. The algorithm is then repeated until the simulation is finished.

\section{Discussion}

\subsection{Interpreting our results}

The two most cited mass loss mechanisms for the bimodal distribution of exoplanet radii are XUV-induced photoevaporation and core powered mass loss. The former mechanism is usually thought to be more consistent with data because it can explain the radius distribution and the sub-Jovian desert, that is, the lack of planets with large radii at very close distances to their host stars \citep[e.g.,][]{Fulton2017,Owen2018}. The core powered mass loss model cannot explain the sub-Jovian desert because it is independent of external heat sources; this has been used as a justification for ignoring the internal heat flux as a source of mass loss. However, the discovery of young small mass exoplanets with densities lower than cold hydrogen \citep[i.e., super-puffs;][]{Benatti2019,Benatti2021,Libby2020} has provided renewed support for the core powered mass loss model because only with high internal luminosities can such enlarged radii be attained \citep{Lopez2014}. In other words, a combination of photoevaporation and core powered mass loss is required for the bimodal distribution, sub-Jovian desert, and super-puff presence to be explained. Our atmospheric evaporation model is the first to unify both mechanisms for estimating the mass loss rate of super-Earth and sub-Neptune exoplanets. This combination gives rise to unforeseen emergent properties, such as the three regimes of atmospheric evaporation. Our findings provide a strikingly new outlook in which photoevaporation and core powered mass loss are not competing explanations but integral parts of the evaporation history of exoplanets.

Regime one applies only to planets with very hot interiors and guarantees an almost total atmospheric loss. Regime two is for planets with cooler interiors when mass loss is still driven by internal energy but the atmosphere has developed a radiative region. Regime three is for planets with even lower temperatures when mass loss occurs primarily from XUV irradiation. Whereas regime three can last for billions of years (evidenced by observations of highly irradiated exoplanets with hydrogen-rich atmospheres), regimes one and two are transient because of the extreme mass loss that occurs as a consequence of high surface temperatures (see the left panel of Figure~\ref{fig:time_evolution}). As shown by Figure~\ref{fig:crossover_mass}, planets with greater masses require higher magma ocean temperatures to be in regime one, which is less likely because radiative cooling takes place between each giant impact \citep[][]{Schubert1986,Sossi2022}. It is therefore more probable for super-Earths and sub-Neptunes with smaller masses to be in regimes one or two than more massive ones, so greater mass loss rates are expected. Not only are more massive planets less prone to total evaporation, but they are also more likely to accrete larger atmospheres \citep[e.g.,][]{Ida2004,Ida2005,Ikoma2012}. This suggests that for very small mass exoplanets, the probability of hosting a primordial atmosphere is close to zero, whereas for exoplanets with large enough masses, the probability approaches one. There must therefore be a critical mass at which half of all planets have primordial atmospheres. Evidence for this critical mass comes from the bimodal distribution of exoplanet radii where the first ($1.3R_{\oplus}$) and second maxima ($2.4R_{\oplus}$) are consistent with planets lacking and having primordial atmospheres, respectively \citep[e.g.,][]{Owen2017,Modirrousta2020b}. Mass measurements are unavailable for most planets in the radius distribution, so the critical mass can only be inferred indirectly through modeling. For example, the first maximum at $1.3R_{\oplus}$ and the minimum point at $1.75R_{\oplus}$ are consistent with planets of mass $3M_{\oplus}$ and $8{-}9M_{\oplus}$ respectively \citep[e.g.,][]{Lehmer2017,Owen2017,Jin2018}. The critical mass must be bound by this range because masses below would not lie in the second maximum, whereas masses above would likely experience runaway gas accretion \citep[e.g.,][]{Stevenson1982,Ikoma2000}. Further constraints on this critical mass are crucial for understanding the evolution of super-Earths and sub-Neptunes. Our model provides a plausible explanation where small mass exoplanets are likely in regime one or two and experience extreme mass loss, and massive planets that just experience regime three can hold onto their primordial atmospheres through geological time. Determining the location of the critical mass requires simulations and will therefore be left for future studies.

\subsection{On the possibility of ice-rich mantles}

It has been suggested that sub-Neptunes may be rich in ices, such as Jupiter's moon Europa and Saturn's moon Enceladus \citep[e.g.,][]{Zeng2019,Mousis2020,Venturini2020}. Our theoretical framework would still apply to planets with such properties, though the numerical values of the parameters would have to be changed. Atmospheric evaporation would still be dominated by hydrogen loss in the early stages of a planet's life because of the thermal decomposition of water into hydrogen and oxygen at the high temperatures after a giant impact. In other words, the atmosphere would experience outflow as prescribed by regimes one and two at high enough magma temperatures. XUV-induced photoevaporation (i.e., regime three) would occur after sufficient cooling if a hydrosphere reforms because water can absorb XUV photons and ionize \citep[e.g.,][]{Johnstone2020}. The fundamental physics of our approach would therefore still apply to planets even if they are ice-rich. 

\subsection{Model limitations}

A major advantage of our model is that it is analytic, so it can provide insights into planetary science, without being obscured by the details of numerical modeling. There are, however, some disadvantages that warrant discussion. In the following, we list the atmospheric and geophysical limitations of our framework.

\subsubsection{Atmospheric model limitations}

Our model assumes an ideal gas with constant opacity, heat capacity ratio, composition, and no chemistry. Real systems are, however, dynamic, as can be seen by the geological history of the terrestrial planets in our solar system. Adopting a more complex numerical prescription, such as an atmospheric general circulation model \citep[e.g.,][]{Leconte2013,Lefevre2021}, wavelength dependent opacities \citep[e.g.,][]{Fortney2007,Nettelmann2011}, and a comprehensive atmosphere-interior chemical model \citep[e.g.,][]{Kite2020,Ito2021} would increase the accuracy of our model, though it would be computationally demanding. As of writing, more detailed atmospheric evaporation simulations do not include the three regimes within their framework because they assume a priori that a thermosphere exists and mass loss is always driven by XUV irradiation \citep[e.g.,][]{Kubyshkina2020}. Our analytic framework provides a stepping-stone for more thorough mass loss models by showing that the evolving thermodynamic conditions of planetary interiors will influence the mass loss mechanism taking place. 

\subsubsection{Geophysical model limitations}
\label{sec:magma_uncertainties}

One reason for our nebulous understanding of the conditions after giant impacts is the poorly constrained equations of state of materials under extreme conditions \citep[e.g.,][]{Jing2008,Jing2011,Lock2017}. In this paper, we used the reference values for the magma ocean conditions of \citet{Solomatov2015} with the specific heat from \citet{Miyazaki2019b}. However, the extent to which these values are valid for very high temperature conditions is uncertain. For example, the viscosity of magma typically follows the empirical Vogel-Fulcher-Tammann law \citep{Dingwell2004}. Extrapolating this law suggests its viscosity could be less than that of water at the extreme conditions after giant impacts. Such low viscosities would influence the magma cooling timescale as well as potentially being outside the domain of the classical 1/3 scaling of convection \citep[e.g.,][]{Iyer2020}. Furthermore, the standard density of the magma ocean used may not be representative of the conditions of all super-Earths and sub-Neptunes because (1) larger mass planets probably have denser magma oceans because of their greater internal pressures and gravities, and (2) the geochemistry of the magma oceans may differ substantially from those of Earth \citep[e.g.,][]{Rouan2011,Modirrousta2021}. Because our model is analytic, it is straightforward to replace the chosen reference values with any future improved data sets.

\section{Conclusion}

In this paper we present our self-consistent atmosphere-interior model that considers internal heat from the cooling nucleus and stellar irradiation (thermal and XUV), for the atmospheric evaporation of super-Earths and sub-Neptunes. We have shown that there are three regimes of atmospheric evaporation that depend on the mass of the nucleus and atmosphere as well as the internal thermodynamic conditions. Regarding the longstanding debate on whether XUV irradiation or core powered mass loss is responsible for the atmospheric evaporation of super-Earths and sub-Neptunes, our theory indicates that they both play important roles but with different timescales. Regimes one and two occur immediately after the last giant impact if the magma ocean temperatures are high enough, when mass loss is efficient but cooling is not because the atmosphere is optically thick. The atmosphere will therefore remain in that regime until the photosphere migrates below the Bondi radius, allowing a thermosphere to form in the upper sections of the atmosphere. Because the thermosphere is optically thin, XUV photons become a major heat source, and the primary mechanism leading to mass loss in regime three. Our model provides a comprehensive framework describing the circumstances at which photoevaporation and core powered mass loss occur. Indeed, the framing of the question should not be whether one mechanism is responsible or the other, but rather when one mechanism is active versus the other.

\section*{Acknowledgements}

This work was sponsored by the US National Aeronautics and Space Administration under Cooperative Agreement {No.\,80NSSC19M0069} issued through the Science Mission Directorate and the National Science Foundation under grant EAR-1753916. This work was also supported in part by the facilities and staff of the Yale University Faculty of Arts and Sciences High Performance Computing Center.

\bibliography{bibliography}{}
\bibliographystyle{aasjournal}

\appendix
\renewcommand{\theequation}{A\arabic{equation}}

\subsection*{Derivation of Jeans' mass loss}

The Jeans' escape particle flux is \citep{Catling2017}
\begin{equation}
    \Phi = \frac{n_{\rm x}}{2^{\frac{3}{2}}\pi^{\frac{1}{2}}}\left(\frac{\mu_{\rm x}}{k_{\rm B}T_{\rm x}}\right)^{\frac{1}{2}}\left(\frac{2GM_{\rm n}}{R_{\rm x}}+\frac{2k_{\rm B}T_{\rm x}}{\mu_{\rm x}}\right)\exp{\left(-\frac{GM_{\rm n}\mu_{\rm x}}{k_{\rm B}T_{\rm x}R_{\rm x}}\right)}.
\label{eq:Jeans_flux}
\end{equation}
To convert equation~\ref{eq:Jeans_flux} to the mass loss rate, the surface area and mean molecular weight are multiplied to give
\begin{equation}
    \dot{M}_{\rm a} = 4 \pi R^{2}_{\rm x} \frac{\rho_{\rm x}}{2^{\frac{3}{2}}\pi^{\frac{1}{2}}}\left(\frac{\mu_{\rm x}}{k_{\rm B}T_{\rm x}}\right)^{\frac{1}{2}}\left(\frac{2GM_{\rm n}}{R_{\rm x}}+\frac{2k_{\rm B}T_{\rm x}}{\mu_{\rm x}}\right)\exp{\left(-\frac{GM_{\rm n}\mu_{\rm x}}{k_{\rm B}T_{\rm x}R_{\rm x}}\right)},
\end{equation}
where the density at the exosphere is found by equating the mean free path of a particle with the local scale height
\begin{equation}
    \rho_{\rm x} = \frac{GM_{\rm n}\mu^{2}}{2^{\frac{1}{2}}\pi d^{2} k_{\rm B} T_{\rm x}R^{2}_{\rm x}}.
\end{equation}
Combining the above two equations leads to
\begin{equation}
    \dot{M}_{\rm a} = \frac{2GM_{\rm n}\mu_{\rm x}}{d^{2}}\left(\frac{\mu_{\rm x}}{\pi k_{\rm B}T_{\rm x}}\right)^{\frac{1}{2}}\left(\frac{GM_{\rm n}\mu_{\rm x}}{k_{\rm B}T_{\rm x}R_{\rm x}}+1\right)\exp{\left(-\frac{GM_{\rm n}\mu_{\rm x}}{k_{\rm B}T_{\rm x}R_{\rm x}}\right)}.
\end{equation}

\subsection*{Derivation of adiabatic flow}

The velocity solution for an ideal, isentropic, continous, and compressible gas is found by solving the continuity equations. Consider a gas flowing through a diverging frictionless pipe; the conservation of mass, momentum, and the isentropic equation of state are
\begin{equation}
    \frac{d\rho}{\rho}+\frac{du}{u}+\frac{d\mathcal{A}}{\mathcal{A}}=0,
\end{equation}
\begin{equation}
    \frac{dP}{du} = -\rho u,
\end{equation}
and
\begin{equation}
    \frac{d\rho}{dP} = \frac{\rho}{P} \frac{1}{\gamma},
\end{equation}
where $\mathcal{A}$ is the area, respectively. By combining the momentum and isentropic equations together one arrives at
\begin{equation}
\begin{split}
    \frac{d\rho}{\rho} &= -\frac{\rho}{\gamma P} u du \\
    &= -\frac{u}{c^{2}} du,
\end{split}
\end{equation}
with $c$ being the speed of sound. Inserting the conservation of mass and integrating gives
\begin{equation}
    \frac{u^{2}}{u^{2}_{\rm B}}-\ln{\left(\frac{u^{2}}{u^{2}_{\rm B}}\right)} = 4 \ln{\left(\frac{r}{R_{\rm B}}\right)}+1,
\end{equation}
where the area, $\mathcal{A}$, has been substituted with $4 \pi r^{2}$, and the speed of sound, c, has been replaced with the speed of sound at the Bondi radius, $u_{\rm B}$.

\subsection*{Derivation of hydrostatic optical depth}

The equation for the optical depth is
\begin{equation}
    \frac{d\tau}{dr} = -\rho \kappa_{\rm th},
\end{equation}
which can be multiplied by the equation for hydrostatic equilibrium to give
\begin{equation}
  \frac{d\tau}{dP} = \frac{\kappa_{\rm th}}{g}.
\end{equation}
Evaluating the above equation and inserting the ideal gas equation gives
\begin{equation}
    \tau = \frac{\kappa_{\rm th} \rho k_{\rm B} T r^{2}}{G M_{\rm n} \mu}
\label{eq:tau_derivation}
\end{equation}

\subsection*{Derivation of energy flux from mass loss}

Consider a parcel of gas located at the Bondi radius. The loss of gravitational potential energy in moving the parcel of gas from the Bondi radius to the gravitational extent of the planet is
\begin{equation}
    \Delta E = GM_{\rm n}\Delta M_{\rm a} \left(\frac{1}{R_{\rm B}} - \frac{1}{R_{\rm Hi}}\right),
\end{equation}
where $R_{\rm Hi}$ is the Hill sphere. The Hill sphere is usually very distant, such as for Earth where it is $230R_{\oplus}$ from its center of mass, so it can be ignored in energy calculations. The above equation therefore becomes
\begin{equation}
    \Delta E \simeq \frac{GM_{\rm n}\Delta M_{\rm a}}{R_{\rm B}},
\end{equation}
which can be differentiated with respect to time,
\begin{equation}
    \frac{dE}{dt} \simeq \frac{GM_{\rm n}}{R_{\rm B}} \dot{M}_{\rm a},
\end{equation}
and converted into the heat flux by dividing through by the surface area of the Bondi radius
\begin{equation}
    F_{\rm ml} \simeq \frac{GM_{\rm n}}{4 \pi R^{3}_{\rm B}} \dot{M}_{\rm a},
\end{equation}

\subsection*{Derivation of the photospheric density}

Starting from equation~\ref{eq:tau_derivation} and inserting $\tau {=} 2/3$,
\begin{equation}
    \rho_{\rm p} = \frac{2GM_{\rm n}\bar{\mu}}{3\kappa_{\rm th} k_{\rm B}T_{\rm p}R_{\rm p}^{2}}.
\label{eq:photosphere_not_hydro}
\end{equation}
Equation~\ref{eq:photosphere_not_hydro} is only valid for atmospheres that are in regime three. Equation~\ref{eq:optical_depth_Bondi} can be used to find the photospheric radius and density with the knowledge that $\rho{\in}\mathcal{O}\left(r^{-2}\right)$ if the atmosphere is in regime one or two:
\begin{equation}
    R_{\rm p} \approx \frac{3}{\gamma}\kappa_{\rm th} \rho_{\rm B} R_{\rm B}^{2} 
\end{equation}
and
\begin{equation}
    \rho_{\rm p} \approx \frac{\gamma}{3 \kappa_{\rm th}R_{\rm p}}.
\end{equation}

\pagebreak

\subsection*{Tables}

\begin{longtable}{| c | c |}
\caption{Subscript meanings} \label{tab:subscripts} \\
\hline
\textbf{Subscript} & \textbf{Description} \\
\hline
\hline
0 & Conditions at zero pressure or reference conditions \\
\hline
a & Atmosphere \\
\hline
ab & Top of the atmospheric boundary layer \\
\hline
bol & Bolometric \\
\hline
B & Bondi radius \\
\hline
c & Crystallization \\
\hline
cmb & Metallic core-mantle boundary layer \\
\hline
cr & Referring to the critical Rayleigh number \\
\hline
n & Nucleus \\
\hline
eff & Effective \\
\hline
eq & Referring to the equilibrium temperature of the planet \\
\hline
EUV & Extreme ultraviolet \\
\hline
H & Referring to Hydrogen \\
\hline
ha & Referring to the hard component of the XUV bands \\
\hline
Hi & Referring to the Hill sphere \\
\hline
i & Ionized \\
\hline
li & Liquidus \\
\hline
mb & Bottom of magma ocean boundary layer\\
\hline
ml & Mass loss \\
\hline
mo & Magma ocean (as a whole) \\
\hline
ni & nonionized \\
\hline
p & Photosphere \\
\hline
pl & Planetary mass (or the mass of the nucleus) \\
\hline
rcb & Radiative-convective boundary \\
\hline
s & Magma ocean surface \\
\hline
so & Referring to the soft component of the XUV bands \\
\hline
t & Turbopause \\
\hline
th & Referring to thermal photons \\
\hline
v & Vapor pressure \\
\hline
w & Referring to the outflowing winds \\
\hline
x & Exobase \\
\hline
XUV & X-ray and ultraviolet\\
\hline
$\oplus$ & Earth \\
\hline
$\odot$ & Sun \\
\hline
$\ast$ & Host star \\
\hline
\end{longtable}

\begin{longtable}{| c | c | c | c | c |}
\caption{Parameters and constants} \label{tab:params} \\
\hline
\textbf{Parameter} & \textbf{Description} & \textbf{Value} & \textbf{Units} & \textbf{Source} \\
\hline
\hline
$a$ & Semi-major axis & ... & m & ... \\
\hline
$A$ & Bond albedo & ... & ... & ... \\
\hline
$\angstrom$ & Angstrom & $10^{-10}$ & m & ... \\
\hline
$\mathcal{A}$ & Area & ... & $\rm m^{2}$ & ... \\
\hline
amu & Atomic mass unit & $1.661 {\times} 10^{-27}$ & $\rm kg$ & ... \\
\hline
AU & Astronomical unit & $1.496 {\times} 10^{11}$ & m & ... \\
\hline
$\alpha$ & Volumetric thermal & ... & $\rm K^{-1}$ & ... \\
 & expansion coefficient & & & \\
\hline
 & Volumetric thermal &  &  & \\
$\alpha_{\rm mb}$ & expansion coefficient of & $5{\times}10^{-5}$ & $\rm K^{-1}$ & \citet{Solomatov2015} \\
 & magma ocean & & & \\
 & boundary layer & & & \\
\hline
$C_{1}$ & Conductivity constant one & $4.00{\times}10^{-2}$ & ... & ... \\
\hline
$C_{2}$ & Conductivity constant two & 0.5 & ... & ... \\
\hline
$c_{\rm p}$ & Specific heat at constant & ... & $\rm J~kg^{-1}~K^{-1}$ & ... \\
& pressure & & & \\
\hline
$c_{\rm p,mb}$ & Specific heat of magma & 5000 & $\rm J~kg^{-1}~K^{-1}$ & \citet{Miyazaki2019b} \\
 & ocean boundary layer & & & \\
\hline
$\mathcal{C}$ & Time coefficient & ... & ... & \citet{Locci2018,Locci2019} \\
\hline
$d$ & Kinetic diameter of particle & ... & $\rm m$ & ... \\
\hline
$D$ & Molecular diffusion coefficient & ... & $\rm m^{2}~s^{-1}$ & \citet{Chapman1970} \\
\hline
$E$ & Energy & ... & J & ... \\
\hline
$\varepsilon_{0}$ & Permittivity of free space & $8.854{\times} 10^{-12}$ & $\rm m^{-3}~kg^{-1}~s^{4}~A^{2}$ & ... \\
\hline
$\varepsilon_{\rm r}$ & Relative permittivity for a gas & ${\sim}1$ & ... & ... \\
\hline
$\xi_{1}$ & Scattering mass loss efficiency & $0.5$ & ... & This work \\
\hline
$\xi_{2}$ & Fraction of incident XUV & $0.8$ & ... & \citet{Locci2018,Locci2019} \\
 & irradiance absorbed & & & \\
\hline
$\xi_{3}$ & The fraction of energy & Equation~\ref{eq:efficiency_3} & ... & This work \\
 & not used for ionization & & & \\
\hline
$\xi_{\rm el}$ & Energy limited efficiency & $0.1{-}0.4$ & ... & \citet{Locci2018,Locci2019} \\
\hline
$\xi_{\rm XUV}$ & Total XUV mass loss & Equation~\ref{eq:efficiency_XUV} & ... & This work \\
 & efficiency & & & \\
\hline
$f$ & Mole fraction & ... & ... & ... \\
\hline
$\bar{f}$ & Average frequency of & $4.836{\times}10^{15}$ & $\rm s^{-1}$ & ... \\
 & XUV photons & & & \\
\hline
$F$ & Energy flux & ... & $\rm W~m^{-2}$ & ... \\
\hline
$g$ & Gravitational acceleration & ... & $\rm m~s^{-2}$ & ... \\
\hline
$G$ & Gravitational constant & $\rm 6.674 {\times} 10^{-11}$ & $\rm m^{3}~kg^{-1}~s^{-2}$ & ... \\
\hline
$h$ & Planck's constant & $\rm 6.626 {\times} 10^{-34}$ & $\rm m^{2}~kg~s^{-1}$ & ... \\
\hline
$H$ & Enthalpy & ... & J & ... \\
\hline
$\mathcal{H}$ & Scale height & ... & m & ... \\
\hline
$I$ & Ionization energy of & $2.178{\times}10^{-18}$ & J & ... \\
 & atomic hydrogen & & & \\
\hline
$k$ & Thermal conductivity & ... & $\rm W~m^{-1}~K^{-1}$ & ... \\
\hline
$k_{mb}$ & Thermal conductivity of the & 2 & $\rm W~m^{-1}~K^{-1}$ & \citet{Lesher2015} \\
 & magma ocean boundary layer & & & \\
\hline
$K$ & Reduction factor & ${\sim}1$ & ... & \citet{Erkaev2007} \\
\hline
$K_{\rm zz}$ & Eddy diffusion coefficient & ... & $\rm m^{2}~s^{-1}$ & \citet{Gierasch1985} \\
 & & & & \citet{Charnay2015} \\
\hline
$k_{\rm B}$ & Boltzmann's constant & $\rm 1.381 {\times} 10^{-23}$ & $\rm m^{2}~kg~s^{-2}~K^{-1}$ & ... \\
\hline
$\kappa_{\rm th}$ & thermal opacity of & $1$ & $\rm m^{2}~kg^{-1}$ & ... \\
 & molecular hydrogen & & & \\
\hline
 & & & & \citet{Cook1964} \\
 $\kappa_{\rm XUV}$ & XUV opacity of & $10^{5}$ & $\rm m^{2}~kg^{-1}$ & \citet{Lee1976} \\
 & molecular hydrogen & & & \citet{Backx1976} \\
 & & & & \citet{Spitzer1978} \\
 & & & & \citet{Chadney2022} \\
\hline
$l$ & Mean free path of particle & ... & m & ... \\
\hline
$L$ & Luminosity or cooling rate & ... & W & ... \\
\hline
$M$ & Mass & ... & kg & ... \\
\hline
$n$ & Particle number density & ... & $\rm m^{-3}$ & ... \\
\hline
$\mathcal{N}$ & Principal quantum number & ... & ... & ... \\
\hline
$\eta$ & Viscosity & ... & $\rm Pa~s$ & ... \\
\hline
$\eta_{\rm mo}$ & Viscosity of magma ocean & 0.1 & $\rm Pa~s$ & \citet{Solomatov2015} \\
\hline
$\sigma$ & Stefan-Boltzmann constant & $5.670{\times}10^{-8}$ & $\rm W~m^{-2}~K^{-4}$ & ... \\
\hline
$P$ & Pressure & ... & Pa & ... \\
\hline
$\rho$ & Density & ... & $\rm kg~m^{-3}$ & ... \\
\hline
$\rho_{\rm mo}$ & Density of magma ocean & $4000$ & $\rm kg~m^{-3}$ & \citet{Solomatov2015} \\
\hline
$Q_{\rm XUV}$ & Incoming XUV luminosity & ... & W & \\
\hline
$q_{\beta}$ & Electron charge & $1.602{\times}10^{-19}$ & C & ... \\
\hline
$R$ or $r$ & Radius & ... & m & ... \\
\hline
$\mathcal{R}$ & Regime & ... & ... & ... \\
\hline
$\rm Ra$ & Rayleigh number & ... & ... & ... \\
\hline
$\rm Ra_{\rm cr}$ & Critical rayleigh number & 1000 & ... & ... \\
\hline
$\mathcal{S}$ & Shape of the XUV bands & ... & $\rm s^{-1}$ & \citet{Locci2018,Locci2019} \\
\hline
$t$ & Time & ... & s & ... \\
\hline
$T$ & Temperature & ... & K & ... \\
\hline
$T_{1}$ & Temperature constant one & $\rm 2.97{\times}10^{4}$ & K & This work \\
\hline
$T_{2}$ & Temperature constant two & $\rm 9.28{\times}10^{4}$ & K & This work \\
\hline
$\tau$ & Optical depth & ... & ... & ... \\
\hline
$\mu$ & Particle mass & ... & kg & ... \\
\hline
$u$ & Average radial wind velocity & ... & $\rm m~s^{-1}$ & ... \\
\hline
$\lambda$ & Mixing length & ... & m & ... \\
\hline
$\gamma$ & Heat capacity ratio & 5/3 & ... & ... \\
\hline
$\gamma_{0}$ & Reference heat capacity ratio & 4/3 & ... & ... \\
\hline
$X$ & Degree of ionization & ... & ... & ... \\
\hline
$Z$ & Atomic number & ... & ... & ... \\
\hline
\end{longtable}

\end{document}